\documentclass[12pt]{article}
\pdfoutput=1

\usepackage{color}
\usepackage{amsmath}
\usepackage{amsfonts}
\usepackage{amssymb}
\usepackage{caption}
\usepackage{graphicx}
\usepackage{slashed}            
\usepackage{subfig}             
\usepackage{xspace}				
\usepackage{cite}

\newcommand{\arXiv}[2]{\href{http://arxiv.org/pdf/#1}{{\tt #2/#1}}}
\newcommand{\arXivold}[1]{\href{http://arxiv.org/pdf/#1}{{\tt #1}}}

\renewcommand{\tilde}{\widetilde} 

\newcommand{\beq}{\begin{equation}}
\newcommand{\eeq}{\end{equation}}

\newcommand{\TeV}{\,\mathrm{TeV}}
\newcommand{\GeV}{\,\mathrm{GeV}}

\newcommand{\bea}{\begin{eqnarray}}
\newcommand{\eea}{\end{eqnarray}}

\newcommand{\eq}[1]{Eq.~(\ref{#1})}

\usepackage[hypertexnames=false]{hyperref}		

\begin{document}
\begin{titlepage}

\vskip.5cm

\begin{center}
{\huge \bf Composite Higgs Sketch} 
\end{center}

\begin{center}
{\bf  {Brando Bellazzini}$^{a,\, b}$, {Csaba Cs\'aki}$^c$, {Jay Hubisz}$^d$, {Javi Serra}$^c$, {John Terning}$^e$} \\
\end{center}
\vskip 8pt

\begin{center}
$^{a}$ {\it  Dipartimento di Fisica, Universit\`a di Padova and INFN, Sezione di Padova,\\Via Marzolo 8, I-35131 Padova, Italy} \\

\vspace*{0.1cm}

$^{b}$ {\it   SISSA,  Via Bonomea 265, I-34136 Trieste, Italy} \\

\vspace*{0.1cm}

$^{c}$ {\it Department of Physics, LEPP, Cornell University, Ithaca, NY 14853}  \\

\vspace*{0.1cm}

$^{d}$ {\it  Department of Physics, Syracuse University, Syracuse, NY  13244} \\

\vspace*{0.1cm}

$^{e}$ {\it Department of Physics, University of California, Davis, CA 95616} \\

\vspace*{0.1cm}

{\tt  
 \href{mailto:brando.bellazzini@pd.infn.it}{brando.bellazzini@pd.infn.it},
\href{mailto:csaki@cornell.edu}{csaki@cornell.edu}, \\
 \href{mailto:jhubisz@phy.syr.edu}{jhubisz@phy.syr.edu},  
 \href{mailto:js993@cornell.edu}{js993@cornell.edu},
 \href{mailto:jterning@gmail.com}{jterning@gmail.com}}
\end{center}

\vglue 0.3truecm

\centerline{\large\bf Abstract}
\begin{quote}

The couplings of a composite Higgs to the standard model fields can deviate substantially from the standard model values. In this case perturbative unitarity  might break down before the scale of compositeness, $\Lambda$, is reached, which would suggest that additional composites should lie well below 
$\Lambda$. In this paper we account for the presence of an additional spin 1 custodial triplet $\rho^{\pm,0}$. We examine the implications of requiring perturbative unitarity up to the scale $\Lambda$ and find that one has to be close to saturating certain unitarity sum rules involving the Higgs and $\rho$ couplings. Given these restrictions on the parameter space we investigate the main phenomenological consequences of the $\rho$'s. We find that they can substantially enhance the $h\to \gamma\gamma$ rate at the LHC even with a reduced Higgs coupling to gauge bosons. The main existing LHC bounds arise from di-boson searches, especially in the experimentally clean channel $\rho^{\pm}\to W^\pm Z\to 3l+\nu$. We find that a large range of interesting parameter space with 700 GeV $\lesssim m_\rho \lesssim $ 2 TeV is currently experimentally viable.

\end{quote}

\end{titlepage}

\newpage



\section{Introduction}
\label{intro}
\setcounter{equation}{0}
\setcounter{footnote}{0}

The major goal of the LHC program is to provide a complete understanding of the mechanism of electroweak symmetry breaking (EWSB). 
From a theoretical point of view, naturalness points either towards supersymmetry or towards strong dynamics as the most promising explanations. The strong dynamics could either directly break the electroweak symmetry as in technicolor/Higgsless models, or produce a composite Higgs which in turn gets a VEV and breaks the symmetry. With the recent hints for a Higgs-like particle near 125 GeV that have emerged from the CMS~\cite{CMSHiggs} and ATLAS~\cite{AtlasHiggs} experiments at CERN, pure technicolor type theories without light scalars are strongly disfavored, however a strongly interacting light composite Higgs~\cite{GeorgiKaplan} can still be a plausible source of EWSB. The main distinction between a weakly coupled Higgs (like in the MSSM) and a composite Higgs arising from strong dynamics is that in the latter case higher dimensional operators do not decouple and will modify the standard model (SM) relations for Higgs couplings (without mixing with other light states). Such modifications to the Higgs couplings from strong dynamics are  particularly important for processes such as electroweak gauge boson scattering, which become non-perturbative at energies of the order of the compositeness scale, $\Lambda_{NDA} = 4 \pi v \simeq 3 \TeV$.

The strong sector is expected to produce other composite bound states besides the Higgs scalar, for example analogs of the $\rho^{\pm,0}$ of QCD. Such states could all be at the cutoff scale $\Lambda$, if the Higgs couplings are 
very close to the SM.
However, if there are sizable deviations of the Higgs couplings due to strong dynamics, then unitarity in $WW$ scattering will break down \cite{unitarity}
before one reaches the cutoff scale. We then expect that some other resonances appear below the cutoff scale to extend the unitarity of the theory up to $\Lambda$. 

In this paper we will assume that the  hints for a 125 GeV resonance do indeed correspond to a Higgs-like particle with couplings similar to the SM values, but with possible ${\cal O}(1)$ deviations due to the strong dynamics, which we  parametrize  with unknown order one coefficients. The Higgs could be significantly lighter than the strong coupling scale because it is a (pseudo-)Nambu-Goldstone boson (pNGB) of an extended global symmetry~\cite{little,MCH}, however this assumption will not play a significant role here. Our parametrization of the effective Lagrangian will include the pNGB Higgs as a special case. 
The requirement of unitarity up to the cutoff scale  will yield relations between the Higgs  parameters and the couplings and masses of the extra composites participating in the scattering amplitudes.

The resonances involved in the unitarization of $WW$ scattering can be classified by their quantum numbers.
We will assume, motivated by the experimental evidence of no large contributions to the $T$-parameter from BSM physics, that the strong sector dynamics preserves a global custodial symmetry~\cite{Sikivie:1980hm,ADMS}.
This is realized (along with EWSB) by the symmetry breaking pattern $SU(2)_L \times SU(2)_R \to SU(2)_{C=L+R}$.
Furthermore, the strong sector will be assumed to be invariant under a custodial parity symmetry, $P_{LR}$, under which the $SU(2)_L$ and $SU(2)_R$ subgroups are exchanged. 
This is motivated by the absence of large modifications of the $Zb_L\bar{b}_L$ vertex from BSM physics \cite{Agashe:2006at}.
Since the longitudinal gauge bosons transform as a $\mathbf{3}$ of $SU(2)_C$, a given resonance to be exchanged in electroweak gauge boson scattering must be contained in $\mathbf{3} \times \mathbf{3} = \mathbf{1} + \mathbf{3} + \mathbf{5}$ of $SU(2)_C$  and  must have positive $P_{LR}$ parity.
In this work, besides the composite Higgs (which is  a singlet of the custodial symmetry), we will
assume the presence of an extra vector resonance in the adjoint of $SU(2)_C$, 
a composite $\rho$.
Such  states are generically present in models of EWSB by strong dynamics. In warped extra dimensional models~\cite{RS} they would  be interpreted as the first KK excitation of the electroweak gauge bosons. A simple concrete example for a model giving rise to the type of effective Lagrangian considered here is the gaugephobic Higgs \cite{Cacciapaglia:2006mz,Galloway:2009xn}, where a bulk Higgs in a warped extra dimension is strongly peaked toward the IR brane with a VEV somewhat larger than the SM value. This suppresses the coupling of the Higgs to the gauge bosons. In such models higher dimensional gauge invariance ensures that no scattering amplitude grows explicitly with powers of energy, giving rise to unitarity sum rules among Higgs and resonance couplings.

The paper is organized as follows. In Section 2 we present the 
low-energy effective theory of a light Higgs and a vector triplet with the most general couplings consistent with a parity symmetry. In Section 3 we show how the additional charged $\rho$'s can enhance the $h\to \gamma\gamma$ decay rate.  We then examine the restrictions unitarity places on the theory in Section 4, and finally present the indirect and collider constraints on the vector resonances in Sections 5 and 6.
Details of the construction of the low-energy effective theory are given in an appendix.



\section{Phenomenological Lagrangian}
\label{sec:phenolag}
\setcounter{equation}{0}
\setcounter{footnote}{0}

To describe the phenomenology of the composite scalar and vector resonances, $h$ and $\rho$, we use the effective Lagrangian approach. 
We review the necessary basic ingredients~\cite{Giudice:2007fh,Contino:2010mh,Contino:2011np} and present the part of the Lagrangian relevant for LHC phenomenology, which includes linear couplings of the Higgs to SM particles and to $\rho^\pm$ and $\rho^0$.
The details of obtaining the full effective Lagrangian based on the CCWZ construction \cite{Coleman:1969sm} approach is presented  in Appendix \ref{CCWZ}.
 
\begin{figure}[!t]
\begin{center}
\includegraphics[width=0.3\hsize]{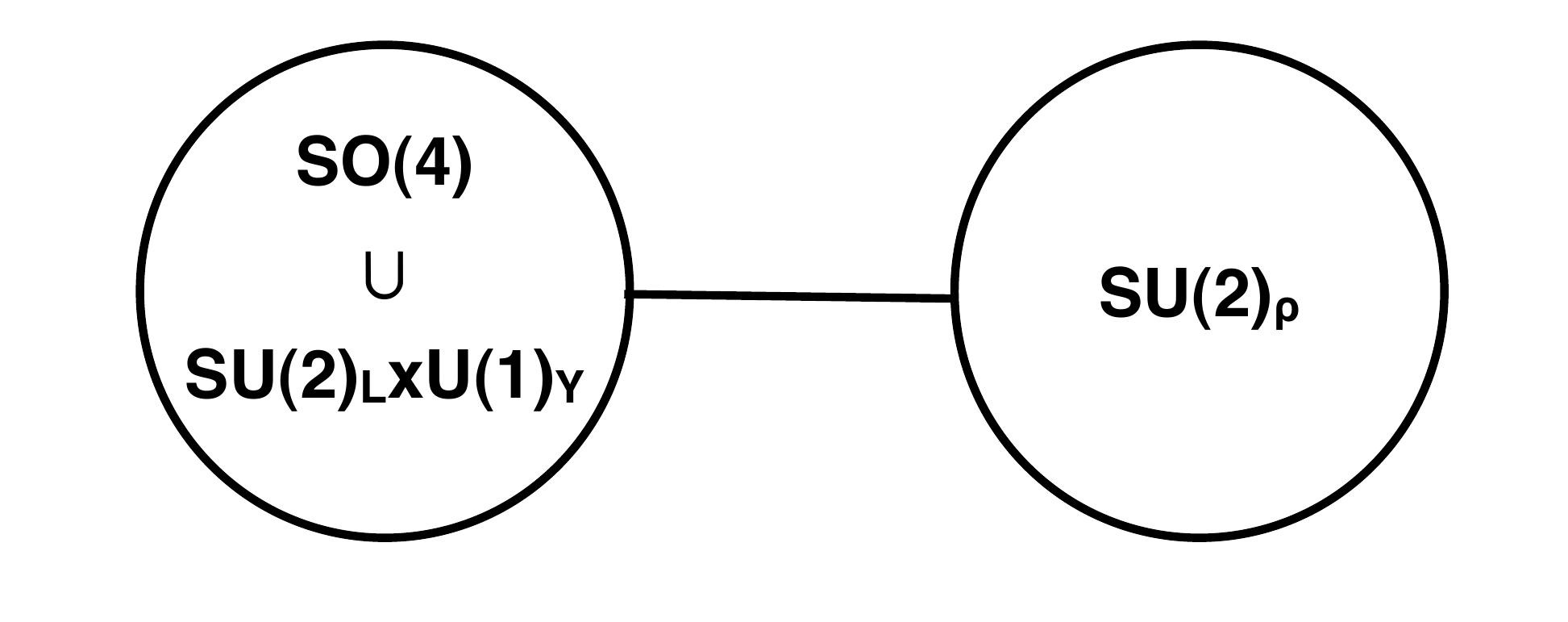}
\caption{The moose diagram for the minimal model considered here. \label{fig:moose1} }
\end{center}
\end{figure}

\begin{figure}[!t]
\begin{center}
\includegraphics[width=0.4\hsize]{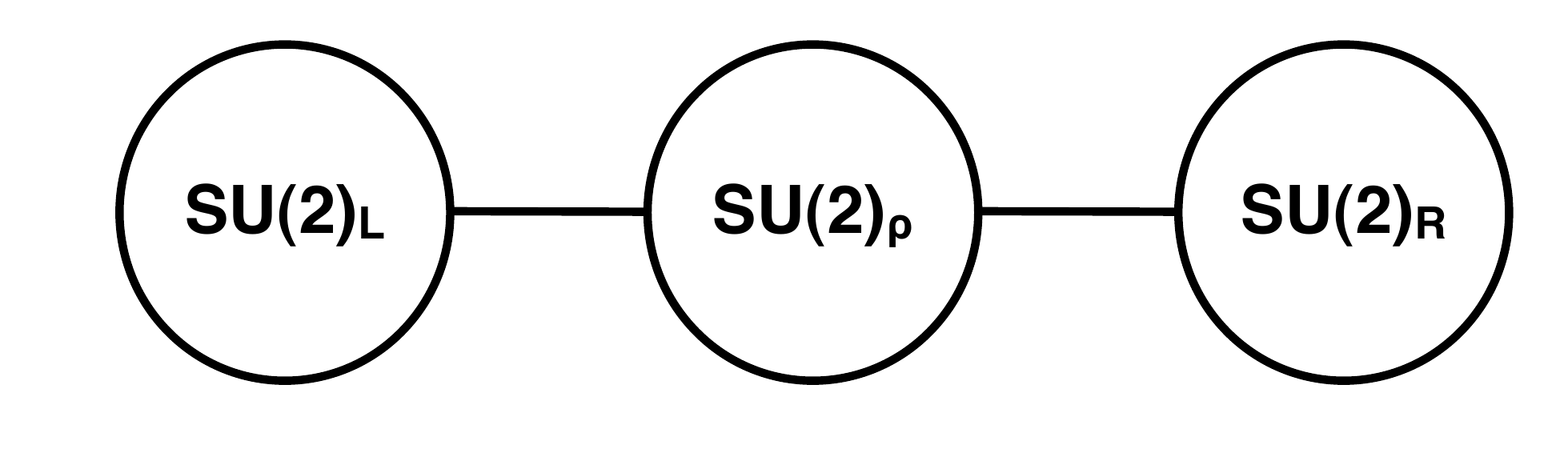}
\caption{A different depiction of the same moose diagram, which has the left-right parity explicit. \label{fig:moose2}}
\end{center}
\end{figure}

We assume that there is a strong sector with an $SU(2)$ custodial symmetry protecting the $T$-parameter, and a (composite) Higgs which is a singlet under the custodial symmetry. Since the coupling of this composite Higgs to the SM gauge bosons may differ from that of an elementary Higgs, we will add to the effective theory  a single spin 1 resonance  $\rho^a$ which is a triplet under custodial $SU(2)$. This triplet is assumed to be lighter than the cutoff scale and its role is to moderate the $VV$ ($V=W^\pm,Z$) scattering amplitudes that  the Higgs fails to fully unitarize up to $\Lambda$ when its couplings deviate from the SM values. The moose diagram corresponding to a minimal effective theory of the sort considered here is given in Fig.~\ref{fig:moose1}. Here the elementary $W^a,B$ spin 1 fields gauge the $SU(2)_L \times U(1)_Y$ subgroup of the global $SO(4)$ symmetry,\footnote{When fermions are included an extra $U(1)_X$ group is needed to get the correct U(1)$_Y$ charges \cite{ADMS}.} and the massive  spin 1 field $\rho^a$ gauges the additional $SU(2)$. The more common depiction of the same moose is given in Fig.~\ref{fig:moose2}, which is obtained by unfolding the two $SU(2)$ factors of the $SO(4)$ global symmetry. Here $SU(2)_L$ is gauged, while only a $U(1)$ subgroup of $SU(2)_R$ is gauged. This notation makes the P$_{LR}$ parity symmetry explicit: it simply corresponds to the exchange of $SU(2)_L \leftrightarrow SU(2)_R$.

\begin{figure}[!t]
\begin{center}
\includegraphics[width=0.4\hsize]{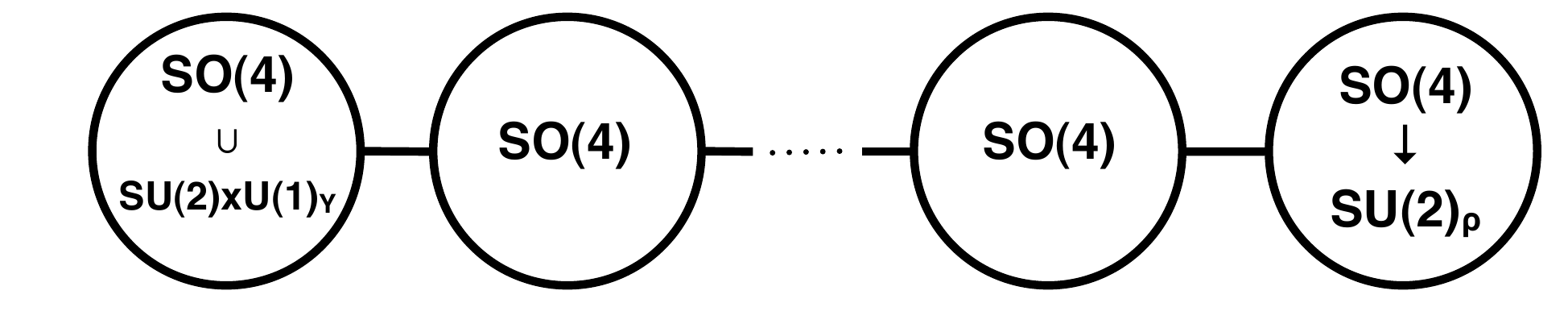}
\caption{The moose corresponding to the deconstructed extra dimensional theory.\label{fig:XDmoose}}
\end{center}
\end{figure}

These moose diagrams can be thought of as a two-site deconstruction~\cite{deconstruction} of an extra dimension with an $SO(4)$ gauge symmetry in the bulk~\cite{ADMS}. The moose diagram of a generic deconstructed theory 
is depicted in Fig.~\ref{fig:XDmoose}, where the last site on the right in the extra dimensional picture would correspond to the $SO(4)\to SU(2)_C$ breaking, making the axial combinations of the $SO(4)$ gauge bosons heavy. Integrating out the axial gauge bosons and limiting ourselves to two sites we obtain the model considered here.  Models with a global symmetry larger than $SO(4)$ will also be described in our formalism, in which case the larger symmetry will impose additional relations among the free parameters considered here. Here we focus on the minimal model, while adding more states/structure will not significantly change our conclusions.

\subsection{Parametrization of the Higgs interactions}

The interactions of the Higgs relevant for its production and decay are parametrized by 
the following effective Lagrangian%
\bea
\mathcal{L}^{(h)}_{\mathrm{eff}}= 
a\left( \frac{2m_W^2}{v}W^+_\mu W^-_\mu+\frac{m_Z^2}{v}Z^2_\mu\right)  h + c_f \left(\frac{m_f}{v} \bar{f} f\right) h + c_\gamma \frac{\alpha}{\pi v} F^2_{\mu \nu} h+c_g \frac{\alpha_s}{12 \pi v} G_{\mu \nu}^2 h \,,
\label{phenolag}
\eea
where the $a$ and $c$'s parametrize possible deviations from the SM Higgs couplings \cite{Contino:2010mh,Carmi:2012yp,Azatov:2012bz,Espinosa:2012ir,Giardino:2012ww} for canonically normalized kinetic terms and masses.  In order to avoid Higgs-mediated flavor changing neutral currents we assume, for simplicity, that the matrix $c_f$ in flavor space is diagonal in the mass basis. The pre-factors have been chosen such that the Lagrangian coincides with that of the SM when all dimensionless parameters are one\footnote{The effective Lagrangian (\ref{phenolag}) already includes one-loop contributions  to $c_{\gamma,g}$ from SM states.}  :
\begin{equation}
a_{\mathrm{SM}} = c_{f,\mathrm{SM}} =1\,, \qquad  c_{g,\mathrm{SM}}\simeq c_{\gamma,\mathrm{SM}}\simeq 1 \,.
\end{equation}
Note that $W^\pm$ and $Z$  have the same $a$ coupling coefficient to the Higgs due to  custodial symmetry. For custodial breaking parametrizations see~\cite{Farina:2012ea}. 
The parametrization in (\ref{phenolag}) is quite general and it captures several models where the Higgs is much lighter than any other state with the same quantum numbers after EWSB.
For instance,  in the MSSM  (or any other two Higgs doublet model with the same structure) one has $a=\sin(\beta-\alpha)$, $c_{t}=\cos\alpha/\sin\beta$, $c_b=-\sin\alpha/\cos\beta$. Models with just another heavy singlet scalar give a universal rescaling $a=c_t=c_b$. Universal rescalings also arise in models where the Higgs is the dilaton of a spontaneously broken CFT \cite{Goldberger:2007zk}, or for a radion in an extra dimension~\cite{radion}.
 In the  scenario where the Higgs is a pNGB the departure of $a$ and $c_f$ from one scales as $v^2/f^2$, with $f$ the Higgs decay constant \cite{Giudice:2007fh}.

\subsection{Effective Lagrangian for the resonance $\rho$}
\label{eff_rho}

Next we consider the interaction terms relevant for the phenomenology of the additional vector resonances $\rho^{\pm ,0}$.  These resonances, together with the isosinglet scalar $h$, can be considered the lowest states in a infinite tower of resonances exchanged in $VV$ scattering. Thus, for this approach to make sense we need the resonance mass $m_\rho$ to be well below the cutoff so that higher dimensional operators induced by extra states (presumably living around the cutoff scale) can be neglected. Under this assumption it is then very convenient\footnote{Although it is not necessary and in some contexts other formalisms are more suitable, see e.g. \cite{Orgogozo:2011kq} and reference therein.}  to represent the resonance as a Lorentz vector $\rho_\mu^a$ that transforms as a gauge field under the unbroken $SU(2)_{c}\subset SU(2)_L \times SU(2)_R$ \cite{Bando:1984ej,Bando:1987br,Georgi:1989xy,Casalbuoni:1985kq}. This will ensure that the gap between $m_\rho$ and $\Lambda$ is naturally stable and the resulting effective field theory is useful because gauge invariance eliminates the  dangerous  operators  suppressed only  by inverse powers of $m_\rho$ rather than the cutoff of the theory. Such operators would be generated by the longitudinal components if the vector were coupled to non-conserved currents. Taking $\rho_\mu$ to be a gauge field renders these longitudinal components harmless because the resonance couples to the conserved current $J^c_\mu$ associated to the custodial symmetry $SU(2)_C$
\begin{equation}
\label{rho_current}
\mathcal{L}_I \supset g_\rho \rho^a_\mu J^{c\, a}_\mu=g_{\rho\pi\pi} \epsilon^{abc} \rho_\mu^a \partial_\mu \pi^b \pi^c+\ldots
\end{equation}
The ellipses in (\ref{rho_current}) contain contributions to $J^c_\mu$ from matter fields as well as from gauge bosons once  the $SU(2)_L\times U(1)_{Y}$ inside $SU(2)_L\times SU(2)_R$ is gauged. These terms give rise to $\rho-W,B$ mixing and therefore, after diagonalization of the mass matrix, generate model-independent contributions to the couplings between $\rho$ and the SM fields.
Additional model-dependent couplings to SM fermions can arise if some chiralities are (partially) composite so that they can couple directly to the $\rho$ via the matter contributions to the current $J^c_\mu$. In the physical mass basis\footnote{We use $\rho^{a=1,2,3}$ for the gauge basis and $\rho^{0, \pm}$ for the physical basis.} the lowest order Lagrangian that is relevant for LHC phenomenology  can then be parametrized  by
\bea
\label{effectiverho}
\mathcal{L}^{(\rho)}_{\mathrm{eff}} \!\!\! &=& \!\!\! c_\rho \dfrac{m_\rho^2}{v}\left(\rho^{0\,2}_\mu +2\rho^+_\mu\rho^{-}_\mu\right)h+ c_{\rho Z} \left(\dfrac{m_Z^2}{v} Z_\mu \rho^{0}_\mu\right) h+  c_{\rho W} \left(\dfrac{m_W^2}{v} W^+_\mu \rho^{-}_\mu+h.c.\right) h 
\nonumber \\
&+&\!\!\! g_{\rho^0 WW}\left(\partial_\mu W_\nu^+ W^{-}_\mu-\partial_\mu W_\nu^- W^{+}_\mu\right) \rho^{0}_\nu +g_{\rho WZ} \left[\left(\partial_\mu W_\nu^- Z_{\mu} -\partial_\mu Z_{\nu}W^-_\mu \right)\rho^{+\nu} + \mathrm{h.c.}\right] + \ldots
\nonumber \\
&+& \!\!\! g_{\rho^0 f} \left(\bar{f} \gamma_\mu T^3_f f\right) \rho^{0}_\mu  + g_{\rho^\pm f} \left(\bar{f} \gamma_\mu T^{\mp} f\right) \rho^{\pm}_\mu \,,
\eea
where the ellipses in Eq.~(\ref{effectiverho}) stands for cyclic permutations of the fields.

\subsubsection{$\rho$ couplings to SM gauge bosons and Higgs}

The first term in (\ref{effectiverho}) is the $h\rho\rho$ coupling which is allowed as long as a mass term for $\rho$ is not forbidden.  Therefore, it is a free parameter which is only constrained by unitarity arguments or NDA.  As explained later, the coupling $c_\rho$ will generate a one-loop contribution to the $h\rightarrow \gamma\gamma$ rate that might be sizable.

The  couplings involving the $\rho$ and the SM gauge bosons $hV\rho$ and  $\rho VV$   are generated from $\rho-V$ mixing after diagonalization of the spin 1 mass  matrix. 
The mixing angle between $\rho$ and $V$ is of order $g_{\mathrm{SM}}/(2g_\rho)$, where $g_{\mathrm{SM}} = g, g'$ are the EW gauge couplings  and $g_\rho$ is the  interaction strength of the vector $\rho$
\beq
\label{trilinear_rho}
\mathcal{L}^{(\rho^3)}= g_\rho\left( \epsilon^{abc}\partial_\mu \rho^a_\nu \rho^b_\mu \rho^c_\nu\right) \, .
\eeq
This latter coupling is assumed to be parametrically larger than $g_{\mathrm{SM}}$, as generically realized in composite models.
After EWSB the resulting $\rho VV$ vertices  are obtained from (\ref{trilinear_rho}) and the SM trilinear $VVV$ vertex. Therefore they have the same Lorentz structure and are of the order $g_{\mathrm{SM}}^2/(2 g_\rho)$,
\begin{align}
\label{rhoVVcoup}
g_{\rho^0 W^+ W^-} \approx - g\left(\frac{g}{4 g_\rho}\right)\,, \qquad 
g_{\rho^\pm W^\pm Z} \approx - g\left(\frac{\sqrt{g^2+g^{\prime 2}}}{4 g_\rho} \right) \,.
\end{align}
The $hV\rho$ couplings are determined by the alignment between the $hVV$ vertex (parametrized by $a$),  and the $h\rho\rho$ vertex (parametrized by $c_\rho$):
\beq
\label{crhoV}
c_{\rho Z} \simeq  (a-c_\rho)\frac{g^2-g'^2}{g_\rho\sqrt{g^2+g^{\prime 2}}}\,, \qquad  c_{\rho W} \simeq (a-c_\rho) \frac{g}{g_\rho} \, .
\eeq
Notice that the couplings of the Higgs to $\rho^0 Z$ vanishes in the limit  $g = g'$.
This is because the $\rho$ is even under the $P_{LR}$ symmetry (which interchanges the $SU(2)_L$ and $SU(2)_R$ groups), while the $Z$ is odd in this limit, so there is no $\rho^0 - Z$ mixing.
The $hV\rho$ vertex  controls the $\rho\rightarrow h V$ decay which has been recently studied using jet substructure techniques \cite{Son:2012mb}. In Section \ref{inelastic_section} we will show that unitarity arguments suggest that $c_\rho=a$, i.e. a vanishing $c_{\rho V}$.

We would like to stress that the Higgs$-$vector system is completely determined by the four parameters $c_\rho$, $g_\rho$, $m_\rho$, and $a$. 
One of the most important points of this work will be to establish a set of (approximate) relations among these parameters so that we can study the LHC phenomenology  by varying just $2$ or $3$ of them, see Section~\ref{Sec_unitarity}. For instance, requiring perturbative unitarity of the scattering amplitudes, in particular $VV$ elastic scattering, we can correlate the parameters 
$a$ and $(g_\rho, m_\rho)$. Unitarity in the  inelastic channels can fix $c_\rho$ as well.  In practice, we are going to make reasonable assumptions based on perturbative unitarity sum rules to reduce the number of free parameters and be more predictive.
In this way we will be able to tie the properties of the Higgs to those of the $\rho$, with important phenomenological implications.
The three parameters that will be used most in the following are
\begin{equation}
\label{param_set}
m_\rho\,,\qquad c_\rho\,, \,\,\,\,\, {\rm and}\,\,\, a\,,
\end{equation}
where $g_\rho$ will be traded for $a$ by imposing the cancellation of the leading term, that grows with $E^2$,  of the $V V$ elastic scattering amplitude. We will impose the unitarity sum rule $a^2+3a_\rho^2/4=1$ where $a_\rho=m_\rho/(g_\rho  v)$. 
More general and thus less predictive conditions could be imposed as well. See e.g. \cite{Contino:2011np} where a more general criterion called \textit{Partial UV completion} fixes only the order of magnitude of $a_\rho$. Our approach differs from \cite{Contino:2011np} because we are assuming only one extra resonance below a relatively high cutoff $\Lambda \sim 3-5$ TeV which will imply that the couplings must be close to their values set by the unitarity sum rules (see Section~\ref{Sec_unitarity}).
 
The choice of the set (\ref{param_set}) is clearly motivated by the fact that $a$ and $c_\rho$ directly correlate with the Higgs partial widths into $\gamma\gamma$ and $WW$
\begin{equation}
\Gamma/\Gamma^{\mathrm{SM}}(h\rightarrow\gamma\gamma)\simeq \left[1+\frac{9}{8} c_\rho +\frac{9}{7}(a-1)-\frac{2}{7}(c_t-1)\right]^2 \,, \qquad 
\Gamma/\Gamma^{\mathrm{SM}} (h\rightarrow WW)=a^2\,,
\end{equation}
so that from these important Higgs decay channels we can immediately learn something about the $\rho$.  

Let us finally notice that in models where the Higgs is a pNGB, the $h\rho\rho$ vertex is forbidden by a shift symmetry, and thus $c_\rho = 0$, up to explicit symmetry breaking effects.
$c_{\rho V}$ still vanishes at leading order in $g_{SM}/g_\rho$, because of $P_{LR}$ symmetry.
However, the larger global symmetry associated to these scenarios, for instance $SO(5)/SO(4)$, implies the presence of extra resonances at low energies, such as an axial vector, to which the pNGB Higgs would couple with generic $\mathcal{O}(1)$ strength, with no $g_{SM}/g_\rho$ suppression.

\subsubsection{$\rho$ couplings to fermions}

We will consider two different physical scenarios for the  
couplings of the $\rho$ to SM fermions,  with distinct phenomenological implications.
In the first scenario, the SM fermions are elementary and the couplings arise only through the mixing of the $\rho$ with SM gauge bosons.  In this case,
\begin{equation}
\label{rho_f_elem}
\mathcal{L}^{\mathrm{el}}_{\rho ff}=-g \left( a_\rho \frac{m_W}{\sqrt{2}m_\rho} \right) \rho^{\pm}_\mu\bar{f}\gamma^\mu T^{\mp}f-\rho^0_\mu \bar{f}\left[\frac{(g^2-g'^2)}{2g_\rho}T^3+\frac{g^{\prime 2}}{2g_\rho} Q\right]\gamma^\mu f \,,
\end{equation}
where 
$T^{\pm}$, $T^3$ and $Q$ are the usual $SU(2)_L$ and $U(1)_Q$ SM charges.
In this case the phenomenology of the $\rho$ is completely determined by the three parameters in (\ref{param_set}). Since in this case there are non-vanishing model independent couplings of the $\rho$ to light quarks, Drell-Yan production of the heavy vectors is always possible. Besides the production cross section, $g_\rho$ and $m_\rho$ also determine the $\rho$ decay widths to gauge bosons and fermions. 

In the second scenario some of the SM fermions are  partially composite. This is due to a mixing
 of the SM chiralities with  fermionic operators in the strong sector.
The most important consequence of this partial compositeness is a direct coupling  between the SM fermions and the $\rho$, in addition to the model independent contribution (\ref{rho_f_elem}). This modifies the $\rho$ production cross section and branching fractions. 
The additional coupling is proportional to the degree of compositeness of a given fermion, $\epsilon_f$, which in principle can be different for each SM chirality.
We can then parametrize the additional  couplings of the $\rho$ to SM fermions as
\beq
g_\rho \epsilon_f^2 \bar{f} \gamma_\mu \tilde{T}^a_f f \rho^{a \mu} \,.
\eeq
It is important to emphasize that these couplings not only depend on $\epsilon_f$, but also on the way the SM fermions couple to the strong sector, which is encoded in $\tilde{T}_f$, as we 
show next in  a particularly relevant example.

To be concrete we consider  a single family of  quarks.  The couplings of the elementary fermions to the strong sector can be written as
\beq
{\cal L}_{\mathrm{mix}}= 
		(\bar u_L, \bar d_L) \epsilon_L^{A} \mathcal{Q}_{A} 	+
		\bar u_R \epsilon_R^{B} \, \mathcal{U}_B +
		h.c.\, ,
\label{mixing}
\eeq
where $A,B$ are $SU(2)_L \times SU(2)_R$ indices, $\mathcal{Q}, \mathcal{U}$ are the composite 
operators\footnote{At low energies these manifest themselves as composite particle excitations.}
and $\epsilon_{L,R}$ parametrize the degree of mixing of the elementary SM states to each of the components of the composite operators.
Of course these interactions are invariant under the SM gauge symmetries, and in particular $SU(2)_L$ invariance implies that the degree of compositeness is the same for $u_L$ and $d_L$.
Furthermore, in order to reproduce the correct hypercharges for the composite fields mixing with the SM fermions, the strong sector must carry extra $U(1)_X$ quantum numbers, such that electric charge is given by $Q = T_L^3 + T_R^3 + X$. 
Flavor and electroweak precision bounds 
suggest then that the composite operators mixing with $q_L$ carry $(\mathbf{2},\mathbf{2})_{2/3}$ quantum numbers.
This allows in particular for the implementation of $P_{LR}$ parity in the coupling of the SM left-handed (LH) bottom, which protects its coupling to the $Z$ from receiving unacceptably large corrections \cite{Agashe:2006at}.
The right-handed (RH) up-quarks can then mix with composite custodial singlets  $\mathcal{U} \sim (\mathbf{1},\mathbf{1})_{2/3}$.
We assume that RH down-quarks also couple to singlets, in this case $\mathcal{D} \sim (\mathbf{1},\mathbf{1})_{-1/3}$.

The representations under which $\mathcal{Q}$, $\mathcal{U}$ and $\mathcal{D}$ transform, along with $\epsilon_{L,R}$, completely determine the low-energy couplings to the $\rho$.
These are obtained following the CCWZ rules, and we again refer the interested reader to Appendix~\ref{CCWZ} for the details.
Since $u_R$ and $d_R$ couple to singlets of $SU(2)_C$, they have no direct couplings to the $\rho$.
Instead, $u_L$ and $d_L$ have the following interactions,
\beq
\label{rho_f_comp}
\mathcal{L}^{\mathrm{co}}_{\rho ff}= c_{\rho f} \frac{g_\rho}{2 \sqrt{2}} \rho^+_\mu \bar{u}_L \gamma^\mu d_L+ h.c. - c_{\rho f} \frac{g_\rho}{2} \rho^0_\mu \bar{d}_L \gamma^\mu d_L \,.
\eeq
Interestingly, the neutral composite vector only couples to down quarks.
This fact can be understood by recalling that we required $P_{LR}$  to be a symmetry of the strong sector, and that the $\rho$'s have positive parity.
Because of the $SU(2)_L \times SU(2)_R$ representation to which we coupled the LH quarks, down quarks have definite parity, so that a $P_{LR}$ preserving coupling to the $\rho^0$ is generated,\footnote{This is the same reason for the absence of a correction to the $Zd_L \bar d_L$ vertex.}
while the couplings of up quarks must arise from a $P_{LR}$ violating coupling.
An observation of $\rho^0$ decays to down and not to up quarks would be an indication of the particular symmetry structure of this strong sector, and of the protection mechanism for the $Z$ couplings to down quarks.

Let us finally make a few remarks on the flavor structure of the strong sector, which is relevant for  the couplings of the $\rho$'s and $h$ to the different SM families.
Some models of particle compositeness predict that SM fermion masses scale with the degree of compositeness as $m_f \propto \epsilon_{L} \epsilon_{R}$.
This, along with the mass hierarchies of the SM quark sector suggest that only the third generation quarks are composite \cite{Pomarol:2008bh}, 
and this will be the assumption we will be using for the couplings of the $\rho$'s.
However, the assumption that the degree of compositeness is universal for LH and/or RH quark families would also be viable.
This scenario is suggested by the tension between the paradigm of partial compositeness and flavor observables like $\epsilon_K$, since it allows for the implementation of Minimal Flavor Violation \cite{Redi:2011zi} or an extra-dimensional GIM mechanism \cite{Cacciapaglia:2007fw}. In this case Drell-Yan production of the vector resonances could be enhanced by a large coupling to light quarks, and there would also be an increased branching fraction to jets.
One must then consider the bounds coming from LHC energetic dijet events \cite{Redi:2011zi,Domenech:2012ai}.


\section{Higgs boson rates: enhancement of $h\to \gamma \gamma$}
\setcounter{equation}{0}
\setcounter{footnote}{0}

\begin{figure}[t!]
\begin{center}
\includegraphics[scale=0.6]{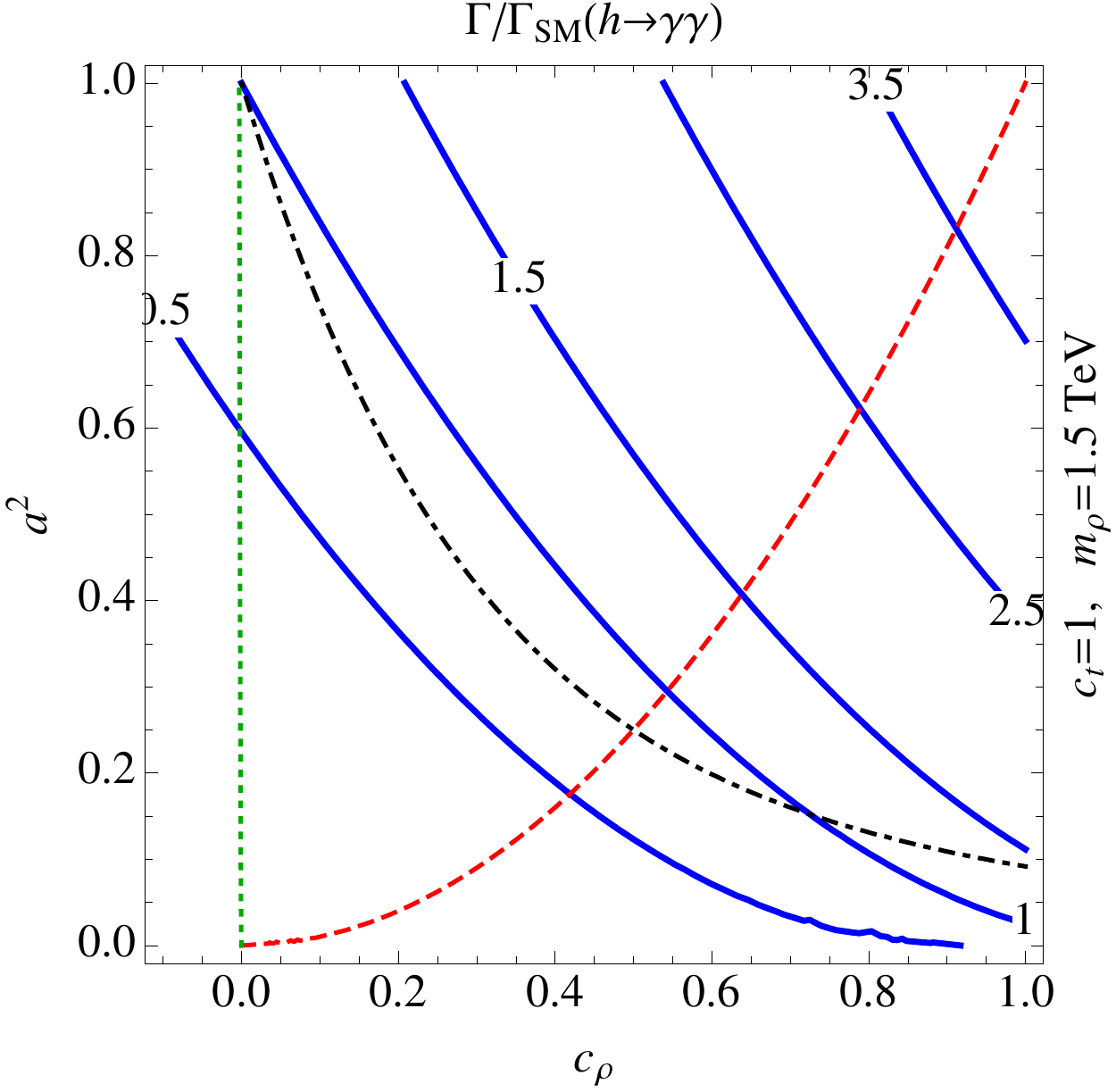}
\hspace{0.5cm}
\includegraphics[scale=0.6]{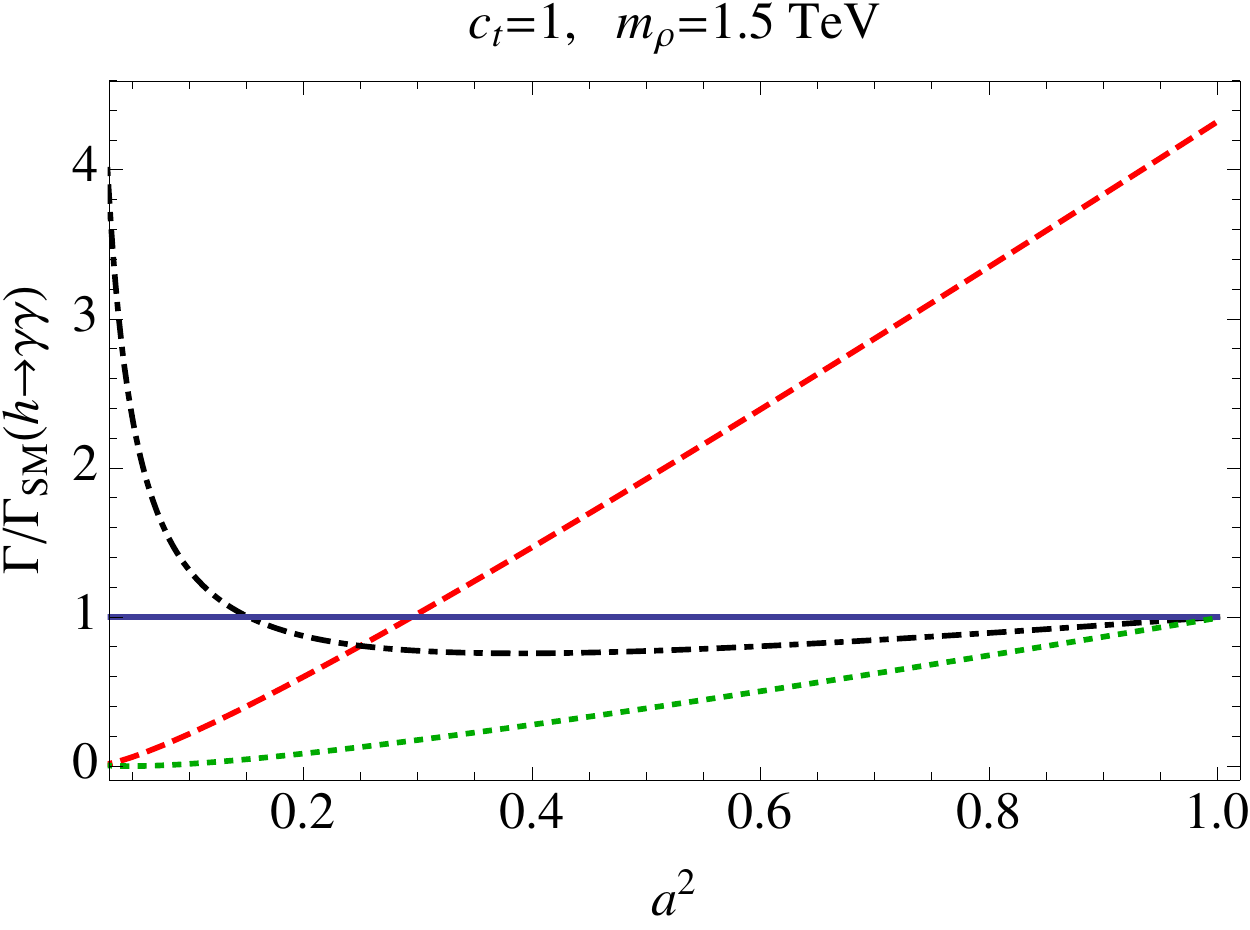}
\caption{$\Gamma/\Gamma_{SM}(h\rightarrow \gamma\gamma)$. Left: contour-lines in the $(c_\rho,a^2)$ plane, for $c_t = 1$. The dashed red and dot-dashed black lines are for the $c_\rho =a$ and $a \, c_\rho  =a_\rho^2/4$ sum rules respectively (that remove the $\mathcal{O}(s)$ growth in $\pi\pi\rightarrow \rho_L h$ and $\pi\pi\rightarrow \rho_L\rho_L$ scattering amplitudes). The dotted green line is for the case of natural little Higgs-like $\rho$.
Right: for $c_\rho$ fixed by the previous sum rules, and $c_t = 1$.
We have taken $m_\rho = 1.5 \TeV$ and $a>0$.
}
\label{gammagamma}
\end{center}
\end{figure}

\begin{figure}[htb]
\begin{center}
\includegraphics[scale=0.6]{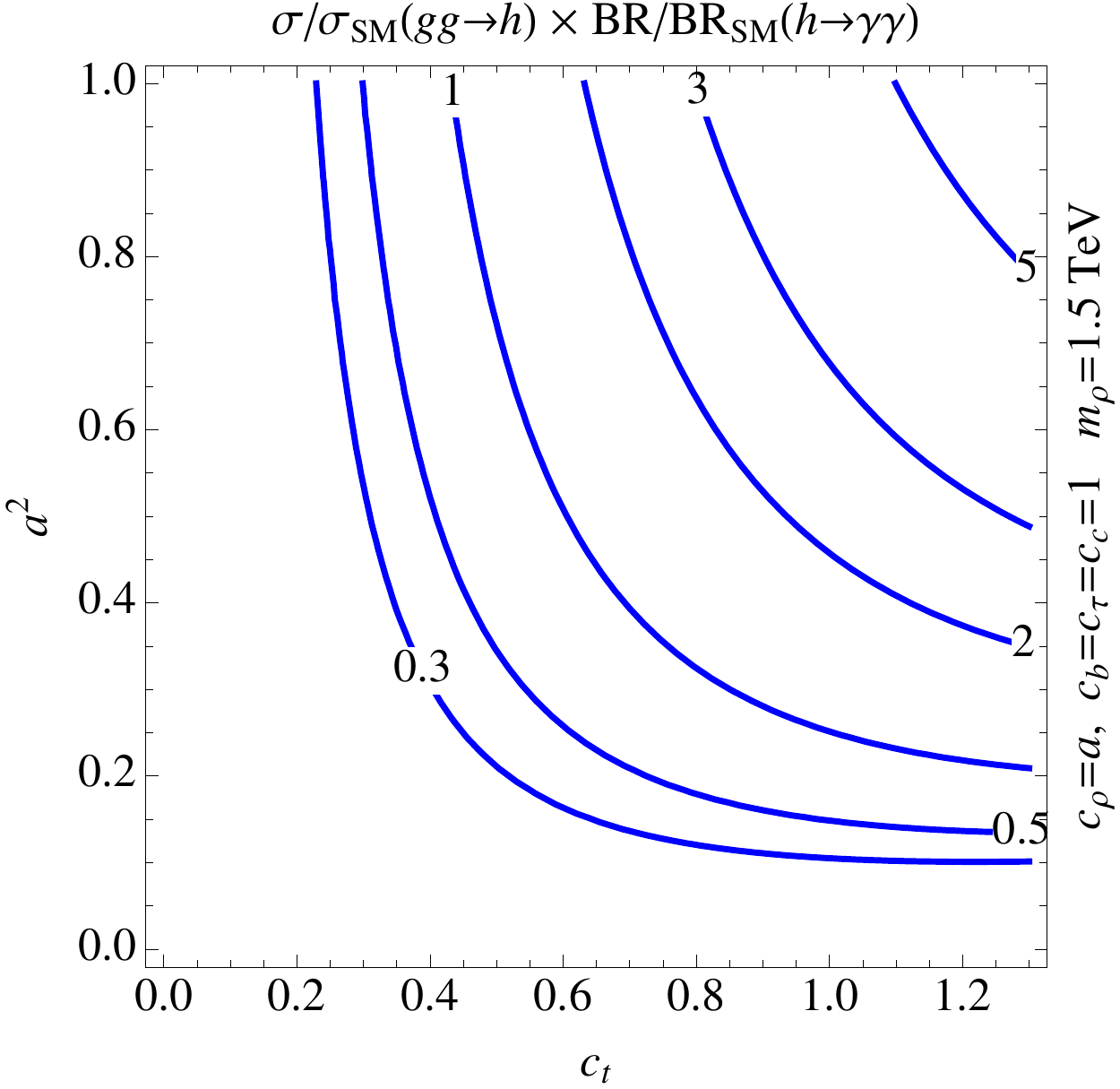}
\hspace{0.5cm}
\includegraphics[scale=0.6]{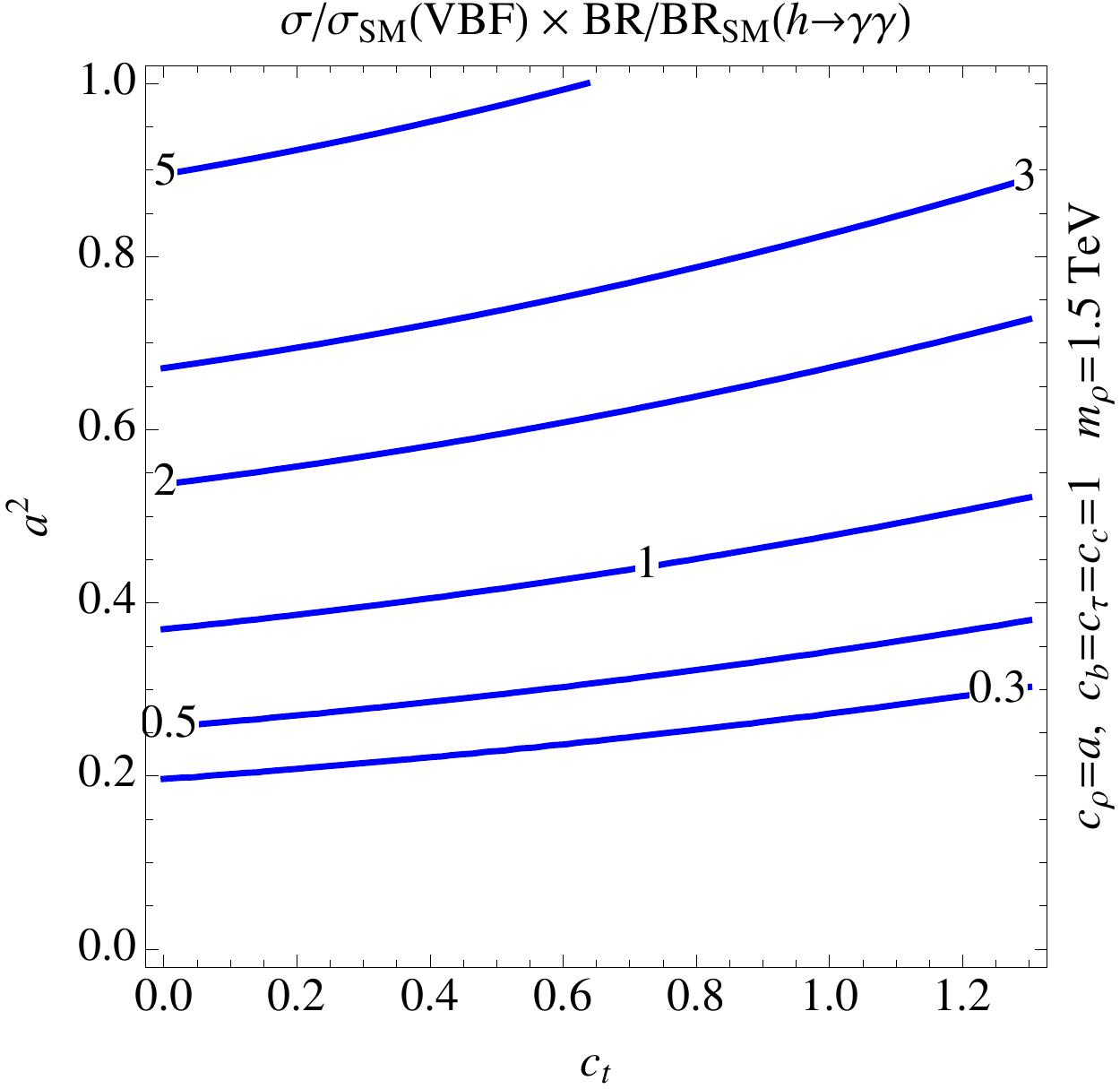}
\caption{Contour-lines for $\sigma \times \mathrm{BR}$ relative to SM prediction in the $(c_t, a^2)$ plane, for Higgs decaying to photon pairs in the gluon fusion channel $gg\rightarrow h \rightarrow \gamma\gamma$ (left) and the vector boson fusion channel $q \bar q \rightarrow \gamma \gamma jj $ (right).
The sum rule $c_\rho = a$ has been enforced, and $c_b, c_\tau, c_c =1$.
We have taken $m_\rho = 1.5 \TeV$ and $a>0$.
}
\label{gammagamma2}
\end{center}
\end{figure}

The Higgs boson effective coupling to photons,  parametrized by $c_\gamma$ in (\ref{phenolag}), is very important at the LHC because it corresponds to  a discovery channel  with small background and good energy resolution, which allows a precise measurement of the Higgs mass. 
This coupling, although vanishing at tree level, is generated at one loop, being sensitive to new physics that contains charged states coupled to the Higgs boson.
In our effective theory this vertex is controlled by the Higgs couplings to $W^+ W^-$, $t\bar{t}$ and $\rho^+ \rho^-$, via $a$, $c_t$ and $c_\rho$ respectively.
The one-loop contribution from these states gives
\begin{equation}
c_\gamma= 
\frac{1}{8}\left[c_t\times N_c \times (2/3)^2 \times F_{1/2}(x_t)+a\times F_1(x_W)+c_\rho \times  F_1(x_\rho)\right]\,,
\end{equation}
where $x_i = 4 m_i^2 / m_h^2$, and 
the values of the functions\footnote{The exact one-loop result used in Figs.~\ref{gammagamma} and \ref{gammagamma2} is given (for $x>1$) by
\begin{align*}
F_{1/2}(x)=-2x[1+(1-x)\arcsin^2(x^{-1/2})]\qquad F_1(x)=2+3 x+3x (2-x)\arcsin^2(x^{-1/2})
\end{align*}
}
$F_{1/2\,,\,1}(x)$ are close to their large $x$ limit for the top and the $\rho$, 
$F_{1/2}(x_t)\approx-4/3$  and $F_1( x_\rho)\approx 7$, 
whereas for the $W$ the contribution is somewhat larger, 
$F_{1}(x_W)\approx 8$. 
The resulting width into photon pairs is thus modified with respect to the SM value,
\begin{equation}
\Gamma/\Gamma_{\mathrm{SM}}(h\rightarrow\gamma\gamma)\simeq \left[1+\frac{9}{8} c_\rho +\frac{9}{7}(a-1)-\frac{2}{7}(c_t-1)\right]^2\,. 
\end{equation}
Possible extra axial vector resonances $A_\mu$, coupled to the Higgs via a $c_a m_a^2 A_\mu^2 h/v$ vertex, can be trivially taken into account just by sending $c_\rho\rightarrow c_\rho+c_a$.

As shown in Fig.~\ref{gammagamma}, a sizable enhancement in the decay rate of the Higgs to $\gamma\gamma$ is possible if $c_\rho$ is not tiny, even with $c_t = 1$.
In fact, unitarity sum rules in the inelastic channels $\pi\pi\rightarrow \rho_L h$ and $\pi\pi\rightarrow \rho_L \rho_L$ yield the relations $c_\rho=a$ and $a\,c_\rho=a_\rho^2/4$ respectively (see Section~\ref{inelastic_section} for more details), which imply important deviations in $\Gamma(h \to \gamma \gamma)$. 
In particular,  as shown in the right panel of Fig.~\ref{gammagamma}, for $a \simeq c_f = 1$ we can get large enhancements
, ranging from 1.5 to 4 times the SM rate.
The sum rule that arises from $\pi\pi\rightarrow \rho_L h$ gives the largest deviations. 
Another approach would be to instead of imposing the unitarity sum rules for the couplings of the $\rho$'s, to require improved naturalness: the coupling of the $\rho$'s would be determined by requiring that they cancel the one-loop quadratic divergence in the Higgs mass due to $W$ loops, in which case $c_\rho=- (m_W^2/m_\rho^2) a$. In this case one finds suppression of the $h\to 2\gamma$ rate.

In models where the Higgs is a composite pNGB, $c_\rho$ is suppressed due to the associated shift symmetry acting on $h$, that forbids any non-derivative couplings.
In fact, the same protection mechanism for the Higgs mass term generically results in a suppression of the $h F^2$ coupling.
Besides, in the pNGB Higgs models one would expect that the leading corrections to $h\to 2\gamma$ would come from the light top-partners needed to cut off the quadratic divergences associated to top loops. 
However, again the effect of these additional fermions on $h \to gg$ or $h \to 2\gamma$ is generically small, because of the protecting symmetry \cite{Azatov:2011qy,Gillioz:2012se}.
Another well-motivated composite higgs model is the higgs as dilaton/radion scenario. In this case the $h\to 2\gamma$ rate can indeed be easily enhanced, and the mass of the dilaton could perhaps remain light, though some tuning is still required~\cite{chacko,ourdilaton}.

From the other partial widths
\begin{align}
\Gamma/\Gamma_{\mathrm{SM}}(h\rightarrow b\bar{b})=&  c_b^2\,, \qquad
\Gamma/\Gamma_{\mathrm{SM}}(h\rightarrow VV^*)=  a^2\,, \qquad
\Gamma/\Gamma_{\mathrm{SM}}(h\rightarrow \tau\bar{\tau}) = c_\tau^2 \,,  \\
\nonumber
 \Gamma/\Gamma_{\mathrm{SM}}(h\rightarrow c\bar{c})=&  c_c^2\,, \qquad
\Gamma/\Gamma_{\mathrm{SM}}(h\rightarrow gg)=  c_t^2\,,
\end{align}
 one can easily calculate the corresponding branching ratios. For instance, the BR into photon pairs is
\begin{align}
&\frac{\mathrm{BR}}{\mathrm{BR}_{\mathrm{SM}}}(h\rightarrow \gamma\gamma)\simeq  \frac{\left[1+\frac{9}{8} c_\rho +\frac{9}{7}(a-1)-\frac{2}{7}(c_t-1)\right]^2}{ c_b^2\mathrm{BR}_{\mathrm{SM}}(h\rightarrow b\bar{b})+a^2 \mathrm{BR}_{\mathrm{SM}}(h\rightarrow VV^*)+\ldots} \,. 
\end{align}
Note, that the apparent non-decoupling effect of the $\rho$-contribution is merely an artifact of the parametrization of the $h\rho\rho$ coupling in~(\ref{effectiverho}). The standard decoupling limit corresponds to $c_\rho m_\rho^2=$ fixed. 

The various Higgs production channels in the SM are also rescaled
\begin{equation}
\frac{\sigma}{\sigma_{\mathrm{SM}}}(gg \to h)\simeq \frac{\sigma}{\sigma_{\mathrm{SM}}}(gg \to h t \bar t)= c_t^2 \,, \qquad 
\frac{\sigma}{\sigma_{\mathrm{SM}}}(q \bar q \to h jj)= \frac{\sigma}{\sigma_{\mathrm{SM}}}(q \bar q \to hW)=a^2\,.
\end{equation}
At the LHC only the product $\sigma \times BR$ is measured.
We show in Fig.~\ref{gammagamma2} the combined effect of the Higgs anomalous couplings on production times branching ratio into photons, normalized to the SM prediction, for two particularly interesting channels in light of current LHC results, gluon fusion $gg \to h \to \gamma \gamma$ and vector boson fusion $q \bar q \to h jj \to \gamma \gamma jj$ (VBF).
Notice that enhancements ranging from 1.5 to 3 times the SM prediction are reproduced for moderately large values $a, c_t \gtrsim 0.7$.  Thus we can see that a suppression of the Higgs couplings to gauge bosons does not necessarily lead to a strong suppression of the $h\to\gamma \gamma$ branching ratio as one might expect for gaugephobic models~\cite{Ellis}.

Finally, we stress that there is another interesting Higgs production mechanism  via the decays of the $\rho$'s. 
In our scenario the decay rate of $\rho \to V h$ is generically small, since it is suppressed by $g_{SM}^2/g_\rho^2$ compared to $\rho \to VV$.
An enhancement of the former channel could be due to $P_{LR}$ breaking.
Also, notice that the decay of an axial vector resonance, $A$,  to $h V$ is not forbidden by parity, and it actually dominates since in this case it is the $A \to VV$ decay that is suppressed.
In both cases $\Gamma(\rho\, (A)\rightarrow V h)$ could be as large as $\Gamma(\rho \rightarrow VV)$ (see Appendix~\ref{rhoVh}) and thus a sizable number of boosted Higgs and gauge bosons would be produced in $q\bar{q}\rightarrow \rho\,(A)\rightarrow V h$ with $\sigma(q\bar{q}\rightarrow \rho,A)\sim$~few fb at 1 TeV.


\section{Unitarity in electroweak gauge boson scattering}
\label{Sec_unitarity}
\setcounter{equation}{0}
\setcounter{footnote}{0}


In this section we wish to explore the regions of parameter space of our low-energy effective Lagrangian where perturbative behavior is retained at high energies, $E \gg m_h, m_\rho$.
We focus on the $2 \to 2$ scattering amplitudes involving the longitudinal components of the electroweak gauge bosons, $V_L$, the composite Higgs, $h$, and the longitudinal components of the extra vector resonance, $\rho_L$.
A priori these amplitudes grow as a power with energy.
By requiring the cancelation of the growing terms, we will identify particularly appealing regions for the parameters 
$a$, $a_\rho = m_\rho/(g_\rho  v)$, and $c_\rho$, thereby improving the consistency of our phenomenological Lagrangian at high energies. In practice we will require a set of precise relations among them, even though some deviations can be allowed.  Recall that in the SM, where EWSB is realized by the VEV of a weakly coupled scalar field, the couplings of the Higgs to electroweak gauge bosons and to itself are such that their scattering amplitudes are unitary up to arbitrarily high energies, for sufficiently low values of $m_h$. The opposite extreme is  the electroweak chiral Lagrangian, where $VV$ scattering remains perturbative only up to the cutoff $\Lambda \lesssim \Lambda_{NDA} = 4 \pi v \simeq 3 \TeV$
(the actual scale is usually well below 2 TeV). 
$\Lambda$ should be regarded then as the scale below which new degrees of freedom should be present.
In Higgsless models \cite{higgsless} the addition of a vector $\rho$ (as considered in this work), can push the value of $\Lambda$ up to $4 \TeV$, while keeping a reasonable hierarchy $\Lambda \gtrsim 2 m_\rho$, for certain values of $m_\rho$ and $g_\rho$ \cite{Falkowski:2011ua}.
In this section we perform 
an analysis of the interplay of Higgs and vector in the unitarization of scattering amplitudes
(see also \cite{Hernandez:2010iu} and \cite{Foadi:2008xj}).
This is of relevance if the couplings of the Higgs measured at the LHC turn out to be different than in the SM, in particular if $a < 1$ in \eq{phenolag}.
We will show how the addition of the Higgs postpones the onset of the vector resonance to higher masses compared to Higgsless models, thus improving compatibility with experimental constraints (see Section \ref{ewpt}).\\

As long as we are only interested in the scattering of longitudinal electroweak gauge bosons, we can make use of the equivalence theorem \cite{Chanowitz:1985hj} to describe their high energy physics with the NGB's $\pi^{a}$ eaten by the $W$ and $Z$, of $SU(2)_L \times SU(2)_R / SU(2)_C \cong SO(4)/SO(3)$.
This is exact up to order $m_W/E$ corrections, thus we will be taking the limit $g, g' \to 0$.
The Lagrangian describing the NGB dynamics can be systematically constructed following the CCWZ method, as reviewed in Appendix \ref{CCWZ}.
Regarding the vector $\rho$'s, although we study amplitudes with their longitudinal components as final states only, we want to consider the exchange of both transverse and longitudinal components in such processes (and besides we want $m_\rho \neq 0$), thus we will keep $g_\rho \neq 0$ and work in a unitary gauge for the $\rho$'s. \\

Before proceeding to the analysis of the scattering amplitudes, let us briefly recall our set-up and the assumptions that go into it.
The electroweak NGB's, $\pi^{a}$, transform linearly under the unbroken custodial symmetry as a triplet.
Our low-energy Lagrangian also contains, as states parametrically lighter than the cutoff, a vector in the triplet representation, $\rho_\mu^a$, and a light scalar singlet, $h$.
The condition $m_\rho, m_h \ll \Lambda$ is a requirement for the validity of the results presented in this section,
since the relations we will be deriving from unitarity are based on a leading order Lagrangian in number of derivatives.
Operators with extra derivatives will be suppressed by powers of $\Lambda$, 
thus a hierarchy between masses and cutoff must be respected.
In practice we will require $\Lambda > (2-3) m_\rho$ (and we consider $m_h < m_\rho$).
One additional assumption is the invariance of the strong sector under the $P_{LR}$ parity (whose action corresponds to the interchange of the $SU(2)_L$ and $SU(2)_R$ symmetry groups).
Under $P_{LR}$, $\pi^{a} \to - \pi^{a}$ while $\rho^{a} \to +\rho^{a}$, and we take $h \to + h$.\\

Finally, in order to systematically estimate the regime of validity of our effective theory, 
we organize the scattering amplitudes in partial waves and $SU(2)_C$ isospin quantum numbers,
\beq
a_{I,J}^{(\alpha,\beta)} = \frac{1}{32 \pi} \int^{+1}_{-1} d(\cos \theta) T^{(\alpha,\beta)}_I(s,\cos \theta) P_J(\cos \theta) \,,
\label{eigen}
\eeq
where $T^{\alpha, \beta}_I$ is the scattering amplitude for the process $\alpha \to \beta$ where the two particle states $\alpha$ and $\beta$ have definite isospin $I$, 
and $P_J$ is the Legendre polynomial for total angular momentum $J$.
A factor $1/\sqrt{2}$ is to be added to account for identical particles in the initial or final state.
Then our condition for perturbative unitarity, which takes into account elastic and inelastic channels, is defined as \cite{Papucci:2004ip}
\beq
\sigma_\alpha | a_{I,J}^{(\alpha,\alpha)} | + \frac{1}{| a_{I,J}^{(\alpha,\alpha)} |} \sum_{\beta \neq \alpha} \sigma_\beta | a_{I,J}^{(\alpha,\beta)} |^2 \leqslant 1 \,.
\label{unitcond}
\eeq
We should be aware of the fact that this condition is somewhat arbitrary, since it fixes the scale where perturbativity is lost. At this scale, by definition, loop corrections introduce order one corrections to the tree-level amplitudes, 
and higher-dimensional operators in a $E/\Lambda$ expansion become important.
For this reason we will consider conservative values of the cutoff $\Lambda$.

\subsection{$\pi \pi$ elastic scattering}

As prescribed by the equivalence theorem, at high energies the elastic scattering of the longitudinal polarization of the $W$ and $Z$ is well described by the NGB's $\pi$.
Their relevant interactions for this process are,
\bea
\pi^4 \!\!\!&:& \left( 1-\frac{3 a_\rho^2}{4} \right) \frac{1}{6 v^2} \left[ (\partial_\mu \pi^{a} \pi^{b})^2-(\partial_\mu \pi^{ a} \pi^{a})^2 \right] 
\label{pi4}\\
\rho \pi^2 \!\!\!&:& \frac{a_\rho^2 g_\rho}{2} \epsilon^{a  b  c} \rho_\mu^{a} (\partial^\mu \pi^{ b}) \pi^{c} 
\label{rhopi2}\\
h \pi^2 \!\!\!&:& 
\frac{a}{v} h (\partial_\mu \pi^{a})^2 \,. \label{hpi2}
\eea
where the indices $a= 1,2,3$ and we recall that $m_\rho = a_\rho g_\rho v$.
Using $SU(2)_C$ invariance and crossing symmetry the amplitude for $\pi \pi$ elastic scattering can be written as
\beq
\mathcal{A}(\pi^{a} \pi^{b} \to \pi^{c} \pi^{d}) = A_s^{(\pi \pi)} \delta^{a b} \delta^{c d} + A_t^{(\pi \pi)} \delta^{ a  c} \delta^{ b d} + A_u^{(\pi \pi)} \delta^{a d} \delta^{b c} \,,
\label{pipi}
\eeq
where $A_s^{(\pi \pi)} = A(s,t,u)^{(\pi \pi)}$, $A_t^{(\pi \pi)} = A(t,s,u)^{(\pi \pi)}$, $A_u^{(\pi \pi)} = A(u,t,s)^{(\pi \pi)}$ is a function of the Mandelstam variables $s$, $t$ 
and $u$ ($s + t + u = 0$). 
One then has
\bea
A(s,t,u)^{(\pi \pi)} &=& \frac{s}{v^2} - \frac{a_\rho^2}{4 v^2} \left[ 3 s + m_\rho^2 \left( \frac{s-u}{t-m_\rho^2} + \frac{s-t}{u-m_\rho^2} \right) \right] - 
\frac{a^2}{v^2} \left[ \frac{s^2}{s-m_h^2} \right] \,,
\eea
where we have not included any widths in the propagators, but these can be introduced trivially.\footnote{Their effect would be relevant for our analysis if the mass of the exchanged particle was close to $\Lambda$ in comparison to the width.
However, in that case higher order operators, suppressed by $(m_h,m_\rho)/\Lambda$ should also be considered.}
The decomposition of such a $\mathbf{3} \times \mathbf{3}$ isospin amplitude in eigenstates of isospin, $\mathbf{1},  \mathbf{3},  \mathbf{5}$, is given by the combinations
\beq
T_0 = 3 A_s + A_t + A_u, \quad T_1 = A_t - A_u, \quad T_2 = A_t + A_u \,.
\label{33}
\eeq

\begin{figure}[!t]
\begin{center}
\includegraphics[width=2.8in]{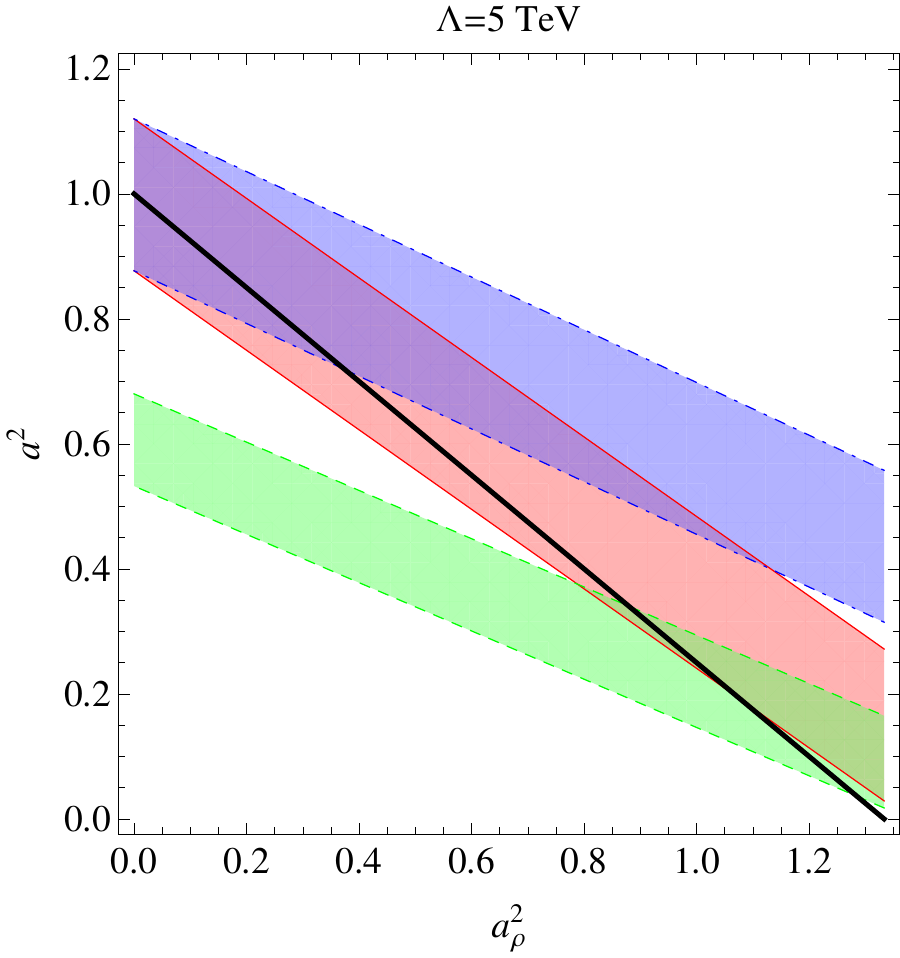}
\hspace{5mm}
\includegraphics[width=2.8in]{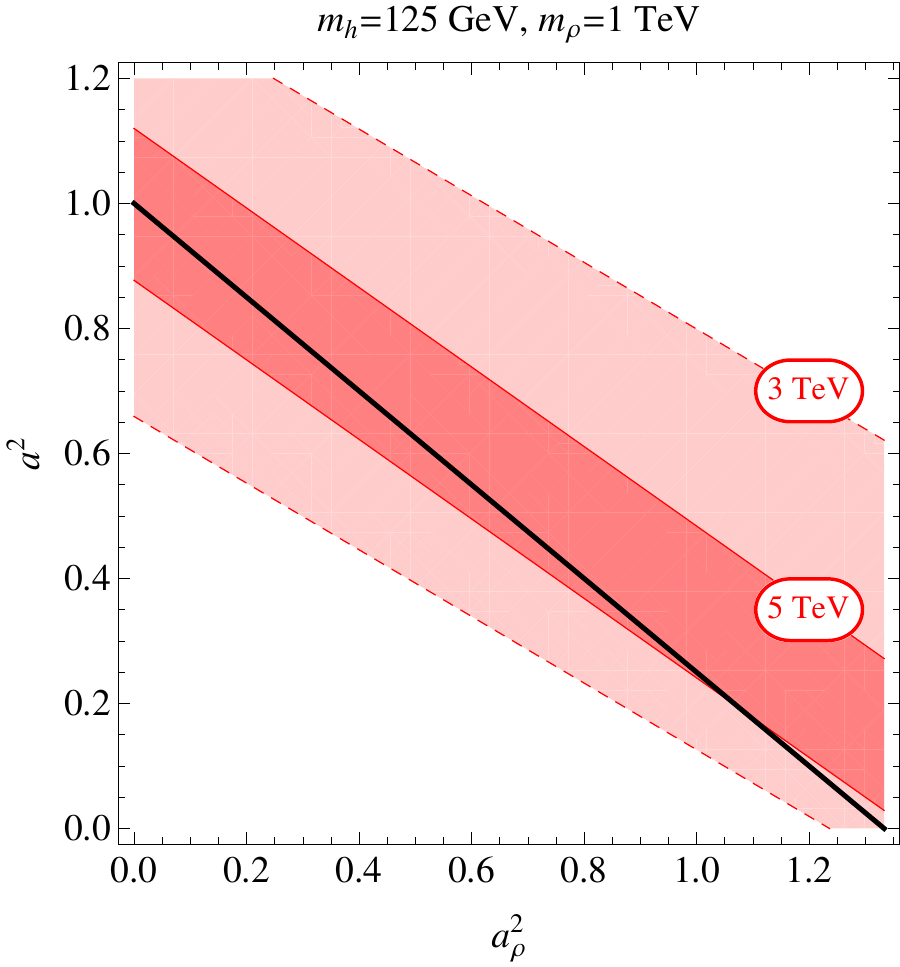}
\caption{Regions allowed by perturbative unitarity in $a_{0,0}^{\pi\pi, \pi\pi}$ ($\pi \pi$ elastic scattering in the isosinglet $s$-wave channel) in the $(a_\rho^2,a^2)$ plane.
Left: requiring a cutoff $\Lambda = 5 \TeV$, for three different mass choices:
$m_h = 125 \GeV$ and $m_\rho = 1 \TeV$ (solid red), $m_h = 2.5 \TeV$ and $m_\rho = 1 \TeV$ (dashed green), and $m_h = 125 \GeV$ and $m_\rho = 2.5 \TeV$ (dot-dashed blue). 
Right: for $m_h = 125 \GeV$ and $m_\rho = 1 \TeV$, and a  cut-off $\Lambda = 5 \TeV$ (solid), and $\Lambda = 3 \TeV$ (dashed). 
The solid black line corresponds to the the sum rule \eq{srpipi}.}
\label{arhoah}
\end{center}
\end{figure}

Regardless of the isospin or angular momentum of the amplitudes for $\pi \pi$ elastic scattering, the linear growth in $s$ for $s \gg  m_\rho, m_h$ will always be cancelled if the following sum rule is satisfied,
\beq
a^2 + \frac{3}{4} a_\rho^2 = 1 \,.
\label{srpipi}
\eeq
Imposing such a relation does not mean that $\pi \pi$ elastic scattering will remain perturbative up to arbitrarily high energies.
The amplitudes still contain logarithmically growing pieces associated with $\rho$ exchange, and finite terms dependent on $m_\rho$ and $m_h$, which may spoil perturbativity.
To explain the importance of these considerations, we show in the left plot of Fig.~\ref{arhoah} the regions allowed by perturbative unitarity in $a_{0,0}^{\pi\pi, \pi\pi}$ ($\pi \pi$ elastic scattering in isosinglet and $s$-wave channel) up to a cutoff $\Lambda = 5 \TeV$ 
in the 
$(a_\rho^2,a^2)$ plane for three different scenarios:
light resonances, $m_h, m_\rho \ll \Lambda$
, a heavy Higgs, $m_\rho < m_h < \Lambda$
, and a heavy rho, $m_h < m_\rho < \Lambda$
, and compared them with the sum rule \eq{srpipi}. 
As expected, the sum rule is most closely followed when both scalar and vector are light.
We also reproduce previous results in Higgsless models \cite{higgsless,Barbieri:2008cc}, where it was shown that in the absence of a scalar, 
$a = 0$, a vector resonance unitarizes most efficiently for values somewhat above the sum rule value $a_\rho = 2/\sqrt{3}$. %
When the scalar is heavy, its relevance in $\pi \pi$ scattering is limited, 
$a \lesssim 0.8$. 
If instead  the vector resonance is heavy, it must be strongly coupled to efficiently unitarize.
In the right plot of Fig.~\ref{arhoah} we show how much one can relax the sum rule \eq{srpipi} when the cutoff is varied from $5$ to $3 \TeV$.
These plots give an idea on how much one can depart from \eq{srpipi} in $\pi \pi$ elastic scattering.
Since we are mostly interested in the scenario with a light $h$ and a relatively light $\rho$,  we are going to assume in the following that the sum rule holds.

\begin{figure}[!t]
\begin{center}
\includegraphics[width=3in]{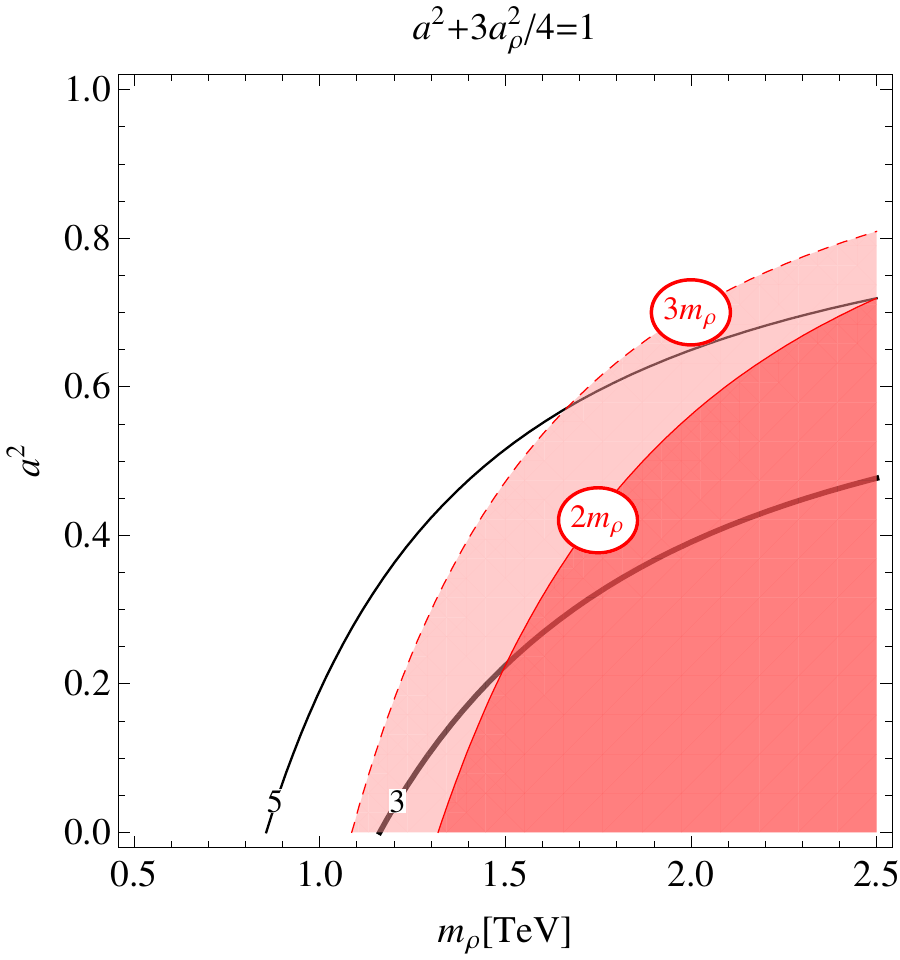}
\caption{Regions excluded by perturbativity unitarity in $a_{0,0}^{\pi\pi, \pi\pi}$ ($\pi \pi$ elastic scattering in isosinglet and $s$-wave channel) in the $(m_\rho,a^2)$ plane, for $\Lambda = 2 m_\rho$ (solid) and $\Lambda = 3 m_\rho$ (dashed).
We also show the exclusion lines for $\Lambda = 3,5 \TeV$, black thick and thin respectively.}
\label{mrhoah}
\end{center}
\end{figure}

If perturbativity of $\pi \pi$ scattering were to give the only non-trivial constraint, in particular from the largest amplitude $a_{0,0}^{\pi\pi, \pi\pi}$, then we could already establish the allowed parameter space in the phenomenologically  interesting plane 
$(m_\rho,a^2)$.
This is shown in Fig.~\ref{mrhoah}, where we consider a light Higgs, $m_h = 125 \GeV$, and we take the cut-off where unitarity is eventually lost at $\Lambda = 2 m_\rho, 3 m_\rho$.
We also show the exclusion lines for $\Lambda = 3,5 \TeV$.
The region 
$a \simeq 1$ is allowed regardless of $m_\rho$, since there the unitarization is carried out almost exclusively by the light Higgs.
For instance, if $\Lambda = 2 m_\rho$ 
then perturbativity admits $\rho$'s as heavy as $2.5 \TeV$ if 
$a \simeq 0.8$.
For smaller values of 
$a$, the unitarization is partly taken over by the vector resonances, which must then be lighter.
Notice that for 
$a = 0$ the heaviest allowed resonance is about $1.3 \TeV$ for $\Lambda = 2 m_\rho$.
In \cite{Orgogozo:2011kq,Barbieri:2008cc,Falkowski:2011ua} heavier vector resonances were allowed since the upper bound $a_\rho \leqslant 2/\sqrt{3}$ was not imposed.
Relaxing our sum rule 
$a^2 + 3 a_\rho^2/4 = 1$ we can relax our upper bound on $m_\rho$ for any given $a$. \\

The next section is devoted to study $\pi \pi$ inelastic scatterings. 
These are relevant to establish further relations in the parameter space of our $h-\rho$ system.
In addition, we will see that the maximum cutoff is actually set by the  inelastic channels in some regions of parameter space, assuming that no additional resonances besides the $\rho$'s appear below 
  $\Lambda \gtrsim 2 m_\rho$.

\subsection{$\pi \pi$ inelastic scattering}
\label{inelastic_section}

We have shown in the previous section that if the coupling of the Higgs to $WW$ deviates significantly from its SM value, the rho's can cure the high energy behavior of the $\pi \pi$ elastic amplitude
, thus keeping a relatively large cutoff compared to the EW scale.
While  a single singlet scalar is enough to fully unitarize a theory with massive $W$'s up to arbitrary high energy scales, this is not the case in the presence of the extra set of massive vectors. The $\rho$ scattering amplitudes put further constraints on the parameter space of our effective Lagrangian.
The relevance of such processes depends on the scale of perturbative unitarity violation:  for instance inelastic $\pi \pi$ scattering into $\rho \rho$ only matters for $\Lambda > 2 m_\rho$.
Recall that new states at or below $\Lambda$ are expected to participate in the unitarization processes.
For instance, in Higgsless models \cite{higgsless} realized in a warped extra-dimension, additional light vector resonances raise the non-perturbative scale above the naive 4D cutoff $\Lambda_{NDA} \simeq 4 \pi v$.
This is accomplished only if the first resonance is light, in which case inelastic channels must also be taken into account.
If instead the resonances are heavy, the 5D cutoff is not much different from the 4D cutoff \cite{Papucci:2004ip}.
This expectation changes in our scenario, thanks to the light Higgs scalar. 
In this case a single extra vector resonance can maintain a moderately large cutoff above $\Lambda_{NDA}$, with masses heavier than in perturbative Higgsless models. 
The scattering amplitudes of $h$ must then also be considered, which give further non-trivial unitarity constraints.

We should however emphasize that additional resonances can contribute to the unitarization of the  $\pi \pi$ inelastic channels, 
without modifying the sum rule, \eq{srpipi}, from the elastic process.
The reason for this is $P_{LR}$ conservation: any state participating in $\pi \pi \to \pi \pi$ must have positive parity, 
thus states with negative parity might modify (or even completely unitarize) some of the channels considered in this section, without affecting our conclusions in the previous section.
We comment on such states below.

The inelastic channels one can consider are $\pi \pi \to hh, \rho_L \rho_L, \rho_L h$. 
Here we present the isospin structure and the high energy behavior of their amplitudes \cite{Hernandez:2010iu}.
This is what we are mostly interested in if we want to derive relations between the parameters of the Lagrangian which render these amplitudes well behaved at high energies.
Several additional interaction terms become relevant here, 
as explained in Appendix~\ref{CCWZ}
\bea
\rho^3 \!\!\!&:& g_\rho \epsilon^{abc} (\partial_\mu \rho^a_\nu) \rho^b_\mu \rho^c_\nu \\
h^2 \pi^2 \!\!\!&:& 
\frac{b}{2 v^2} (\partial_\mu \pi^{a})^2 h^2\\
h \rho^2 \!\!\!&:& c_\rho a_\rho^2 g_\rho^2 v h (\rho_\mu^a)^2.
\eea
The general structure of the  $\pi \pi \to \rho_L \rho_L$ scattering amplitude is of the form
\beq
\mathcal{A}(\pi^{a} \pi^{b} \to \rho^c_L \rho^d_L) = A_s^{(\rho_L \rho_L)} \delta^{a b} \delta^{c d} + A_t^{(\rho_L \rho_L)} \delta^{a c} \delta^{b d} + A_u^{(\rho_L \rho_L)} \delta^{a d} \delta^{b c} \,,
\eeq
where $A_s^{(\rho_L \rho_L)} = A(s,t,u)^{(\rho_L \rho_L)}$, $A_t^{(\rho_L \rho_L)} = B(s,t,u)^{(\rho_L \rho_L)}$, $A_u^{(\rho_L \rho_L)} = B(s,u,t)^{(\rho_L \rho_L)}$, and
\bea
A(s,t,u)^{(\rho_L \rho_L)} &=& \frac{s}{v^2} \left( 
a \, c_\rho - \frac{a_\rho^2}{4} \right) + \cdots 
\label{Arhorho} \\
B(s,t,u)^{(\rho_L \rho_L)} &=& \frac{s}{4 v^2} (a_\rho^2-1) + \frac{t}{4 v^2} (a_\rho^2-2 ) +
\label{Brhorho}
\cdots
\eea
where the ellipses stand for sub-leading terms in $s, t, u$ ($s+t+u=2m_\rho^2$).\footnote{
We thank Haiying Cai for pointing out an error in \eq{Arhorho} in an earlier version of this draft.
}
This amplitude can again be decomposed in $\mathbf{1},  \mathbf{3},  \mathbf{5}$ eigenstates of isospin, as in \eq{33} for $\pi \pi \to \pi\pi$.

For $\pi \pi \to h h$, an isospin-0 process, one has
\beq
\mathcal{A}(\pi^{a} \pi^{ b} \to hh) = A(s,t,u)^{(hh)} \delta^{ab} \,,
\eeq
with
\beq
A(s,t,u)^{(hh)} = \frac{s}{v^2} 
(a^2 - b) + \cdots
\eeq

For $\pi \pi \to \rho_L h$, an isospin 1 process, one finds
\beq
\mathcal{A}(\pi^{a} \pi^{ b} \to \rho_L^{c}h) = A(s,t,u)^{(\rho_L h)} \epsilon^{abc} \,,
\eeq
where
\beq
A(s,t,u)^{(\rho_L h)} = i \frac{t-u}{2 v^2} 
(a - c_\rho) a_\rho \,.
\eeq
For this process the $s$-wave amplitude vanishes, thus it is the vector channel that gives the strongest unitarity constraint.

From these results, one can draw several conclusions.
The inelastic channel $\pi\pi\to hh$ will not lead to a violation of perturbative unitarity, as long as the Higgs is light, if $b = a^2$.
However, the $\pi \pi \to \rho_L \rho_L$ and $\pi\pi\to \rho_L h$ can not be simultaneously unitarized in all channels without additional states.
The linear growth of \eq{Brhorho} proportional to  $s$ or $t$ cannot be both eliminated (these would show up in different partial wave amplitudes).  The cancellation of the remaining growing terms  in $\pi \pi \to \rho_L h$ and $\pi \pi \to \rho_L \rho_L$ in \eq{Arhorho}, along with the sum rule from $\pi \pi \to \pi \pi$ imply the specific values $a_\rho = 1, 
a = 1/2, c_\rho = 1/2$.
This is easily understood from the fact that the our effective Lagrangian with massive $W$'s, $\rho$'s and the scalar has a ``weak ultraviolet completion" into a two Higgs doublet model if an extra scalar, $H$, odd under $P_{LR}$, is introduced \cite{Hernandez:2010iu} (in addition to the $h$ which is assumed to be even under $P_{LR}$).
The extra coupling of this additional $H$, of the form $a_H v g_\rho H (\partial_\mu \pi^{ a}) \rho^{b \mu} \delta^{ a b}$, contributes to the term growing linearly with $t$ in \eq{Brhorho} (without contributing to the term growing with $s$),  which allows all amplitudes to be simultaneously unitarized for $a_H = 1/2$.
It is important to notice that $H$ would not couple linearly to $\pi \pi$, so it would not modify the behavior of $\pi \pi$ elastic scattering. 
Furthermore, its couplings to fermions could be set to zero, since the unitarization of processes $\pi \pi \to f \bar{f}$ can be carried out completely by the Higgs if $c_f = 1$. %
Therefore unitarization of the $\pi \pi$ inelastic scattering amplitudes could be carried out by extra heavy states without  affecting the LHC phenomenology of the $W$'s, $h$ and $\rho$'s very much.
Another possible state that could partially unitarize $\pi \pi$ inelastic channels without affecting the elastic one is an axial vector resonance, that is, a copy of the $\rho$ which is odd under $P_{LR}$.
Such a state could participate in the unitarization of all the inelastic channels. The introduction of such an axial vector generically gives a negative contribution to the $S$-parameter, which could help alleviate the bounds on $m_\rho$ from electroweak precision bounds (see Section \ref{ewpt}).

We would like to emphasize that the theories considered here are not QCD-like. In QCD the $\rho$'s appear right around the cutoff scale, and thus one does not gain any information about the couplings by considering unitarity arguments. The main constraint that the couplings in QCD follow is that asymptotic freedom manifests itself in the form factors, leading to $a_\rho = \sqrt{2}$, motivating vector meson dominance~\cite{Zohar}.

Following the previous discussion, we will be assuming  that the following sum rules are satisfied
\bea
\label{hh}
&&b = a^2 \,, \\
&&c_\rho = a \,.
\label{hrho}
\eea
They lead to the cancellation of the terms 
growing  with energy  in $\pi \pi \to hh$ and $\pi \pi \to \rho_L h$ amplitudes, 
thus they point to a particularly interesting region of parameter space for the couplings $b$ and $c_\rho$ where the perturbative behavior of our Lagrangian is improved.
They are not modified by the inclusion of extra states such as the parity odd Higgs, or the axial vector resonance as long as its coupling to $(\partial_\mu \pi) h$ is small.
The sum rule of \eq{hh} is also satisfied in the SM, while \eq{hrho} corresponds to the case where the  mass and the coupling to the Higgs of the vector bosons are aligned, as shown in \eq{crhoV}.
Nevertheless, these relations are not  robust predictions, thus one should keep a flexible approach.
Finally notice that amplitudes with different isospin or angular momentum select different components of the full scattering amplitudes, thus some of them are sensitive to the sum rules while others are not.
This is the reason why it is important to consider different channels.

In order to show how effective the sum rules Eqs.~({\ref{pipi}, \ref{hh}, \ref{hrho}) are in maintaining perturbative unitarity of scattering amplitudes, we show in Fig.~
\ref{plotmrho}
the maximum allowed cutoff from the requirement \eq{unitcond}, 
as a function of 
the mass of the $\rho$, for two particular values of the Higgs couplings to $WW$, $a = 0.5$ (left) and $a = 0.8$ (right).
\begin{figure}[!t]
\begin{center}
\includegraphics[width=3.0in]{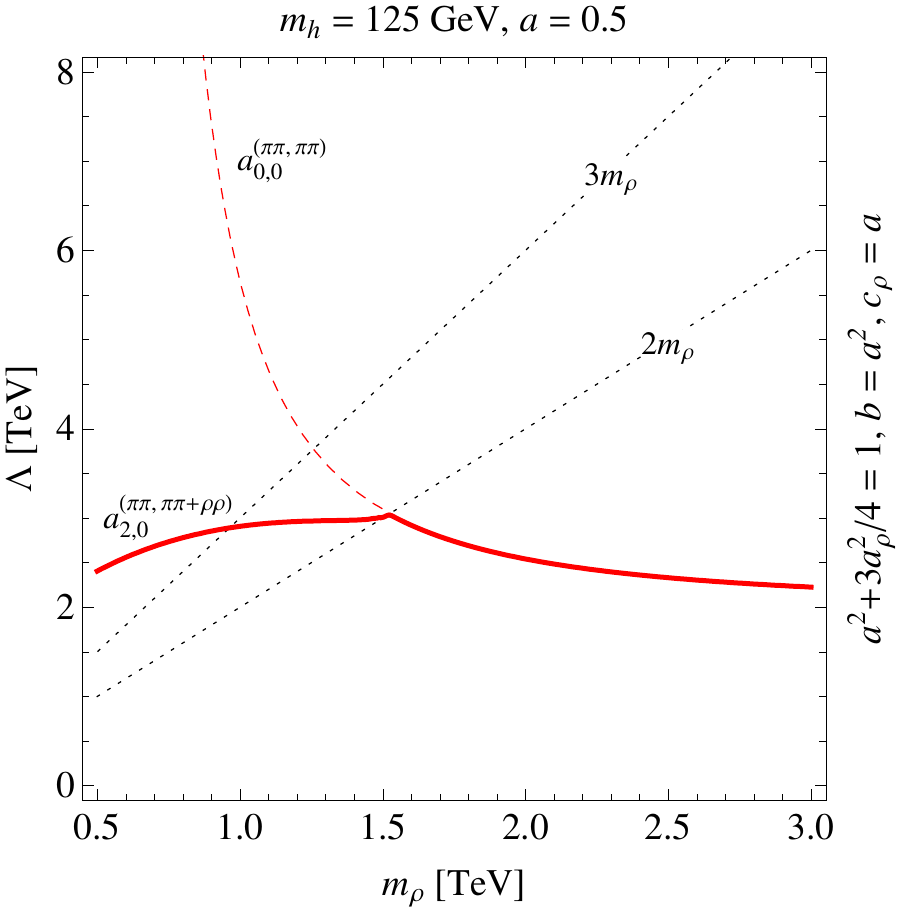}
\hspace{5mm}
\includegraphics[width=3.0in]{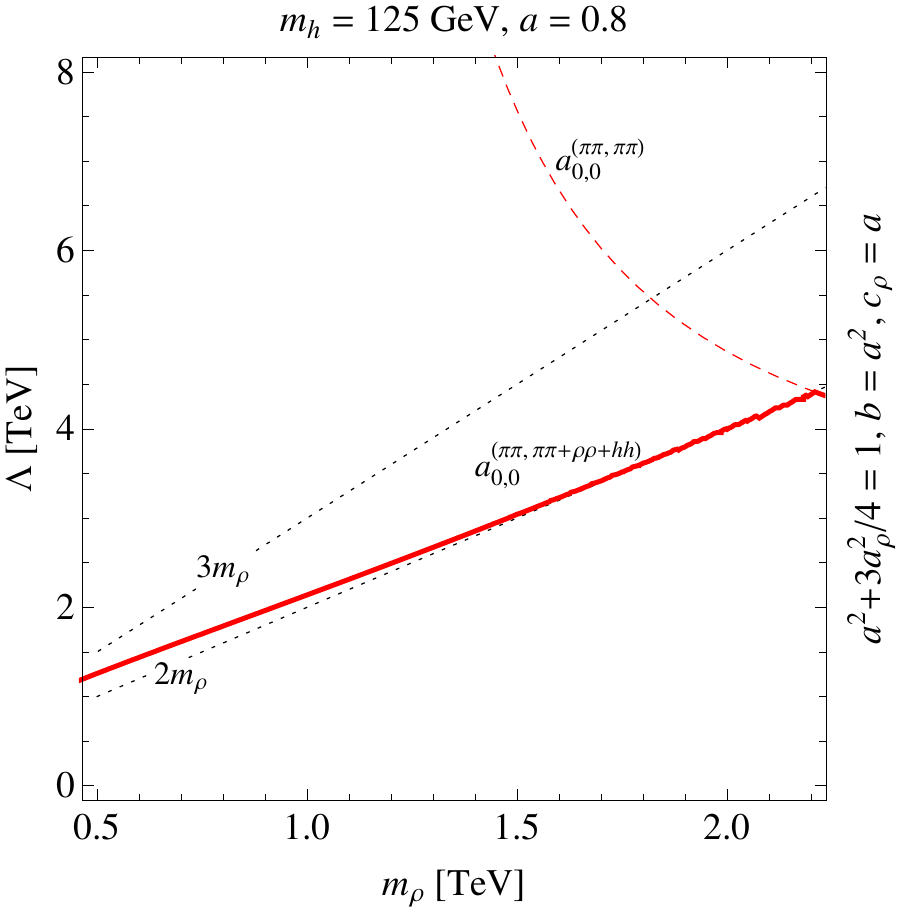}
\caption{The unitarity violation scale $\Lambda [\TeV]$ in the $\pi \pi$ scattering amplitudes versus the mass of the vector resonance $m_\rho$,
for $a = 0.5$ (left) and $a=0.8$ (right).
The dashed line is obtained considering the elastic $\pi \pi$ scattering only, in the isospin-0 s-wave channel, $a_{0,0}^{\pi\pi, \pi\pi}$.
The solid lines include inelastic scatterings, in the isospin-2 s-wave channel $a_{2,0}^{\pi\pi, \pi\pi+\rho \rho}$ (left) and in the isospin-0 s-wave channel $a_{0,0}^{\pi\pi, \pi\pi+\rho \rho+hh}$ (right).
The dotted lines correspond to $\Lambda = 2 m_\rho, 3 m_\rho$.
The maximum value of $m_{\rho}$ shown corresponds to $g_\rho \simeq 4 \pi$.}
\label{plotmrho}
\end{center}
\end{figure}
We recall again that the values of $\Lambda$ obtained from unitarity are not very robust, %
although several important pieces of information can be obtained.
The Higgs scalar significantly improves the perturbative behavior of scattering amplitudes over the Higgsless (or very gaugephobic) case, as can be visualized by comparing the left and right plots of Fig.~\ref{plotmrho},
and recalling that the prediction for a Higgsless scenario is effectively obtained for $a = 0$.
Even though for any given value of $m_\rho$, $\Lambda$ is not much higher than $2 m_\rho$, 
the theory remains under better perturbative control for  quite heavy vector resonances: $m_\rho$ can be above $2 \TeV$ when the Higgs is slightly gaugephobic ($a = 0.8$) and light.
If the Higgs is quite gaugephobic ($a = 0.5$), only light $\rho$'s below $1.5 \TeV$ allow perturbative control of scattering amplitudes while $\Lambda > 2 m_\rho$.
This is problematic because of the associated large contributions to the $S$-parameter.
We understand then that the larger the modification of the Higgs couplings to $W$'s, the lighter the $\rho$'s must be, increasing the tension with EWPT.
Notice further that in those regions where $\Lambda < 2 m_\rho$, it is the $\pi \pi$ elastic scattering that fixes the scale of unitarity loss.
This is mostly due to the fact that  neither $\pi \pi \to hh$ nor $\pi \pi \to \rho_L h$ grow faster than $s$, thanks to enforcement of the associated sum rules.
For $\Lambda > 2 m_\rho$ it is $\pi \pi \to \rho_L \rho_L$ that fixes $\Lambda$.
This changes if for the coupling $c_\rho$ one requires for example the cancelation of the growing terms in \eq{Arhorho}, with $a \, c_\rho=a_\rho^2/4$, rather than imposing \eq{hrho}.  
In this case it is no longer true that $\Lambda < 2 m_\rho$ is fixed by the elastic scattering:  in the region where $\rho$ couples strongly, $a \sim 0$, unitarity is lost in the inelastic channel to $h \rho_L$.

Finally, let us point out that the limit $a \to 1$ does not on its own yield particularly large $\Lambda$'s, and one may wonder why the cutoff does not become very large in this case. The reason is that one still has the inelastic $\pi\pi\to \rho\rho$ channels to worry about, which will be unitarized only if in addition one takes the $\rho$-mass to infinity and thus $m_\rho \gg s$.



\section{Electroweak Precision Constraints} 
\label{ewpt}
\setcounter{equation}{0}
\setcounter{footnote}{0}

Electroweak precision constraints have long been known to plague strongly coupled models of electroweak symmetry breaking.  Due to mixing of vector resonances of the strong sector with electroweak gauge fields, predictions for weak scale observables, such as 
$W$ and $Z$ coupling and masses, 
are modified from their SM values. 
 In this section we study the most dangerous of these, encoded in the $S$ and $T$ parameters, and identify regions of parameter space in which electroweak precision observables do not disfavor strong dynamics.

The electroweak precision constraints 
can be encoded in
the oblique parameters~\cite{oblique1,Barbieri:2004qk,Cacciapaglia:2006pk} if the light quarks and leptons are fundamental. 
By integrating out the $\rho$ triplet we can determine the low-energy corrections to the
transverse self energies $\Pi_{V}(p^2)$ at tree-level where $V=\{W^{+}W^{-}, W_{3}W_{3}, BB, W_3 B\}$:
\begin{align}
\Pi^\prime_{W_3 B}(0)= \frac{1}{4 g_\rho^2} \,, \qquad &\Pi^\prime_{W_3 W_3}(0)=\Pi^\prime_{W^+ W^-}(0)=\frac{1}{g_2^2}+\frac{1}{4g_\rho^2} \,, \qquad \Pi^\prime_{BB}(0)=\frac{1}{g_1^2}+\frac{1}{4g_\rho^2} \nonumber \\
& \Pi_{W^+ W^-}(0)=  \Pi_{W_3 W_3}(0)=-v^2/4\,.
\end{align}
From these expressions one can extract the EWPT parameters $\hat{S}$ and $\hat{T}$ at tree-level:
\begin{equation}
\hat{S}=\frac{g_2^2}{g_2^2+4g_\rho^2}\simeq a_\rho^2 \frac{m_W^2}{m_\rho^2} \,, \qquad \hat{T}=0 \,,
\end{equation}
where we use the notation of~\cite{Barbieri:2004qk}.
The $\hat{T}$-parameter vanishes because of custodial symmetry.
The other oblique parameters, $\hat{U}$, $V$, $X$, $Y$, $W$ and $Z$ are either vanishing or suppressed by extra powers of  $m_W^2/m_\rho^2$ with respect to $\hat{S}$.  Constraints on the model are thus dominated by $\hat{S}$.  In addition to this tree-level contribution from UV resonances, there  are other  sizable contributions \cite{Barbieri:2007bh} from Higgs loops with non-SM couplings to $W$ and $B$ gauge bosons, $a\neq 1$:
\begin{align}
\delta{\hat S}_{\mathrm{IR}}= &\frac{g^2}{96\pi^2 } \left[ \left( 1- a^2 \right) \log \left( \frac{\Lambda}{m_h} \right) + \log \left( \frac{m_h}{m_h (\text{ref})} \right) \right]\\
\delta{\hat T}_{\mathrm{IR}}= &-\frac{3g^{\prime\,2}}{32\pi^2}\left[ \left( 1- a^2 \right) \log \left( \frac{\Lambda}{m_h} \right) + \log \left( \frac{m_h}{m_h (\text{ref})} \right) \right]\, ,
\end{align}
 where we take $m_h (\text{ref}) = 117$~GeV, as in~\cite{Erler:2008zz}. 
 The other UV contribution to $\hat{T}$ from loops with $\rho$ is generically much smaller. Assuming the unitarity sum rule $c_\rho=a$ for simplicity%
\footnote{For $c_\rho \neq a$ we find no extra divergent contributions to \eq{TUV}.} 
one finds \cite{Orgogozo:2011kq,Falkowski:2011ua,Barbieri:2008cc,Torre:2011bv}
\begin{equation}
\delta \hat{T}_{\mathrm{UV}}=\frac{3g^{\prime\,2}}{128\pi^2}a_\rho^2\left[(3-a_\rho^2)\log\frac{\Lambda}{m_\rho}-\frac{1}{3}\log\frac{m_\rho}{m_W}\right]\,.
\label{TUV}
\end{equation}
We show in Fig.~\ref{Sbound} the contours of $\hat{S}=10^{-3}$  and $\hat{T}=-10^{-3}$ on the $(m_\rho, a^2)$ plane,  assuming also a cancellation (tuning) against other contributions, e.g. tree-level axial vector resonance, vertex corrections due to fermion compositeness \cite{Cacciapaglia:2004rb,Cacciapaglia:2006gp}, one-loop contributions from either vectors or fermions,  
and/or higher-dimensional operators \cite{Contino:2011np}.
 For example, it is possible to reduce the strong sector contribution to $\hat{S}$ by adding an axial vector that transforms in the adjoint of $SU(2)_C$ with the following Lagrangian
\begin{equation}
\mathcal{L}_{\mathrm{axial}}=-\frac{1}{4}A^{{a} 2}_{\mu\nu} +\frac{1}{2}m_a^2 A^{ {a} 2}_\mu+\frac{1}{2}\alpha g_a v^2 (gW^{  a}-\delta^{ {3}a} g^\prime B_\mu)A_\mu^{  a}+\ldots
\end{equation}
that in turn gives
\begin{equation}
\hat{S}\rightarrow \hat{S}-\alpha^2 \frac{m_W^2}{m_a^2}\,.
\end{equation}
See \cite{Orgogozo:2011kq} for a more detailed and complete analysis on the spin-1 resonances contribution to $\hat{S}$ and $\hat{T}$.

\begin{figure}[htb]
\begin{center}
\includegraphics[scale=0.6]{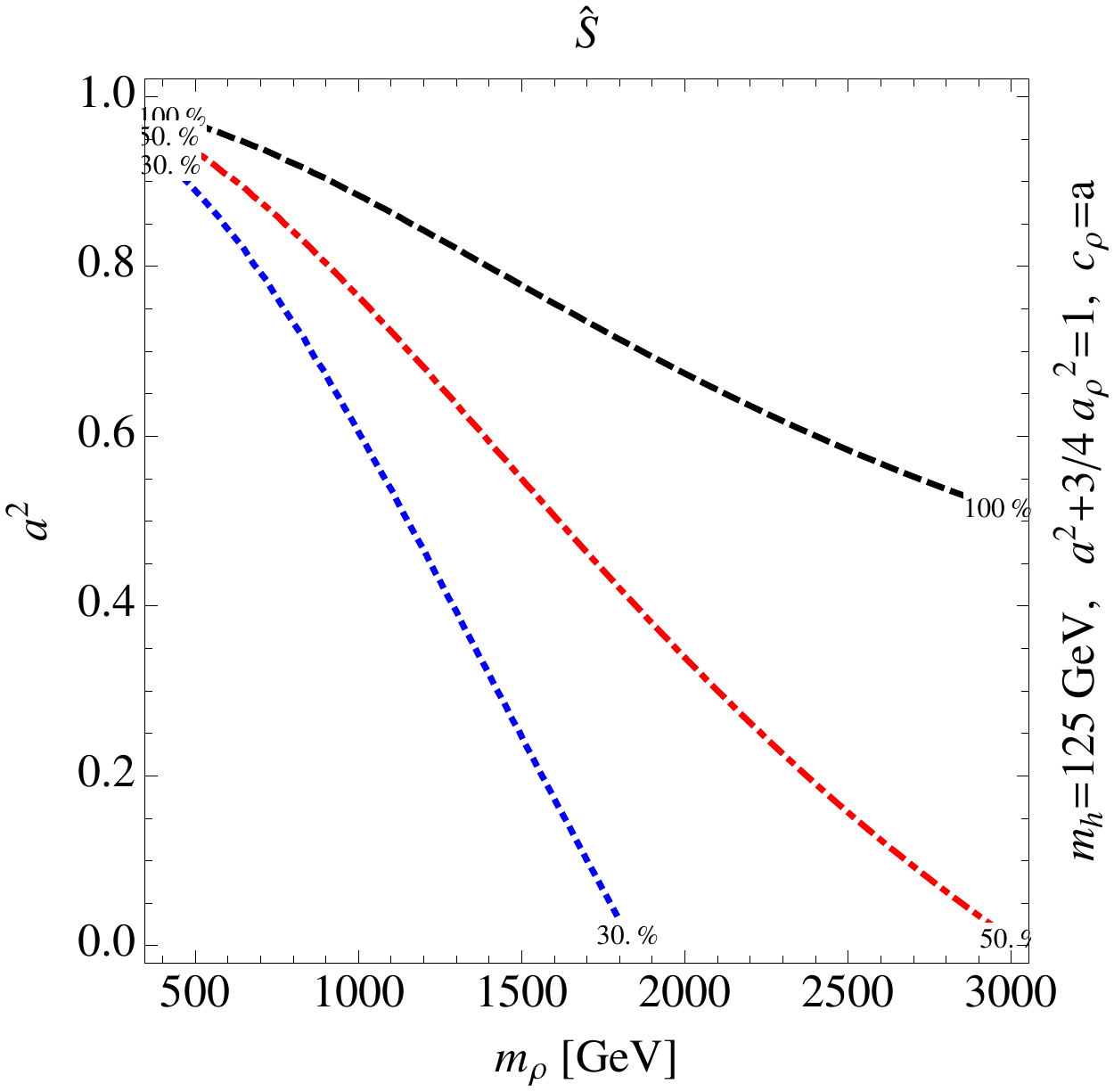}
\includegraphics[scale=0.6]{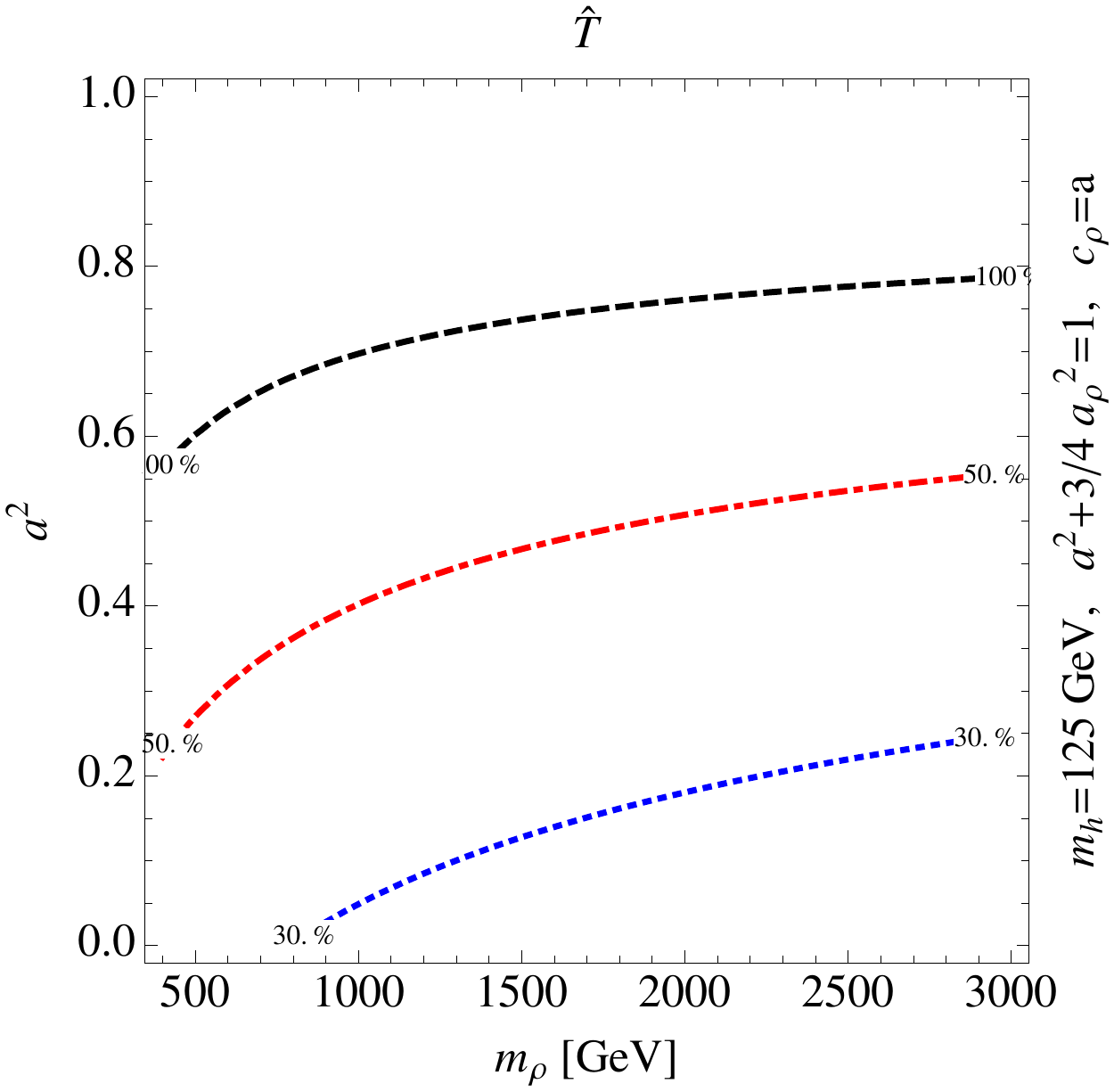}
\caption{Contour-lines for $\hat{S}=10^{-3}$ (left) and $\hat{T}=-10^{-3}$ (right) with $\Lambda=3$ TeV, assuming cancellations  between  the $\rho$ and other contributions of $100\%$, $50\%$ or $30\%$ fine tuning.}
\label{Sbound}
\end{center}
\end{figure}

\section{Direct Collider Constraints from the LHC}
\setcounter{equation}{0}
\setcounter{footnote}{0}

The collider phenomenology of the strongly coupled sector is determined primarily by the mixing of fundamental states with composite ones.  The mixing of the custodial triplet of $\rho$'s with the SM gauge bosons is proportional to  $g_{SM}/g_\rho$.  The SM fermions may also have some amount of compositeness, by mixing with new heavy fermionic degrees of freedom.  These mixings determine contributions to electroweak precisions observables, the LHC production mechanisms, and the relevant decays of the new vector degrees of freedom, all of which are important signals of compositeness at the TeV scale.

Production of resonances in the strong sector proceed through mixing.  There are two production mechanisms for the heavy vector bosons: Drell-Yan and vector-boson-fusion (VBF).  VBF occurs as a result of the mixing between the fundamental gauge bosons and the composite $SU(2)_C$ triplet of vectors.  This mixing leads to couplings of the form $VV\rho$, where $V = W^\pm, Z$.  This same mixing also couples fundamental fermions to the heavy vectors.  Composite fermions may couple directly to the triplet themselves, depending on their $SU(2)_C$ representation.  Light SM fermions (which may be a mixture of fundamental and composite states) couple to the $\rho$'s through a combination of these two channels.

The $\rho-V$ mixing also opens up decays of the heavy vectors to fundamental fermions and electroweak gauge bosons.  Decays of the $\rho$'s  to composite fermions will also contribute substantially to the widths if this is kinematically allowed (i.e. if the SM fermions carry a substantial degree of mixing with composite states).

In this section, we explore the parameter space of the model in light of the LHC searches for di-boson resonances, which place the strongest direct constraint on the allowed masses and couplings of the vector rho of the strong sector.  In the gauge sector, there are two main parameters after taking into account the unitarity constraint from $\pi\pi$ elastic scattering, $a$ and $m_\rho$.  Fixing these two parameters determines $a_\rho$ and $g_\rho$, which in turn fix the vector boson mixing angles in the neutral and charged sectors.  The remaining parameters correspond to the mixings of fundamental and composite states that yield the SM fermions.  Since the full set of mixing parameters is very large,  we will consider only two phenomenologically interesting and theoretically motivated cases:  the case where all LH SM fermions are completely fundamental, and the case where the third generation fermions are completely composite.

To determine the constraints, we have implemented the model in MADGRAPH~\cite{Alwall:2011uj}, and calculated the relevant production cross sections for the charged and neutral heavy vectors using the CTEQ6l1 pdf set~\cite{Pumplin:2002vw}.  The strongest bounds arise from null di-boson resonance searches for $\rho^\pm$ fields at the CMS experiment~\cite{CMSWZres}, as we discuss below.   In Fig.~\ref{fig:rhobound1}, we show the total Drell-Yan cross section for producing the charged $\rho$'s for the cases when $a = 0$ and $0.9$.  The VBF channel has a cross section that is too small to be probed in the current data sets, and requires far higher luminosity (see, e.g.,~\cite{Birkedal:2005yg} and\cite{Belyaev:2008yj}).

\begin{figure}[ht]
\begin{center}
\includegraphics[height=2.5in]{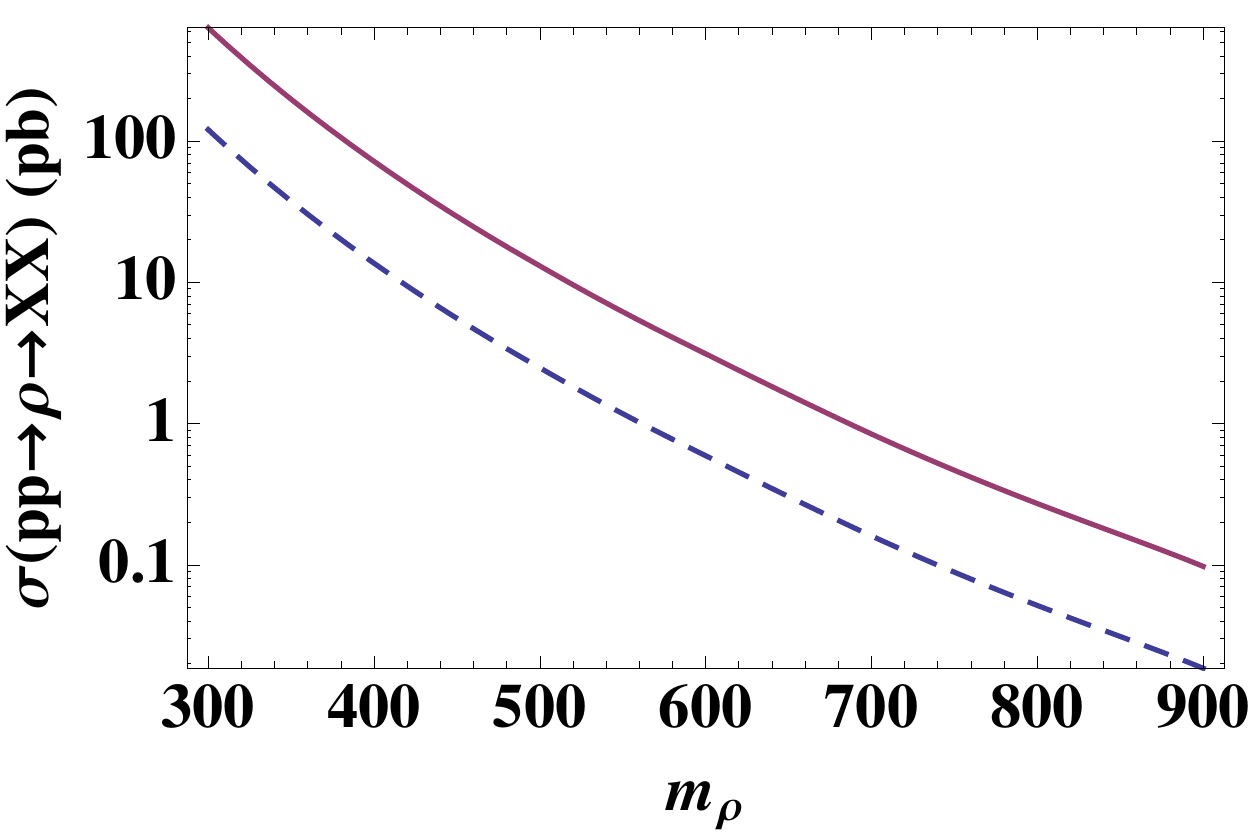}
\caption{The total LHC7 cross section for resonant production of the charged $\rho$'s for $a = 0$ (solid line) and for  $a=0.9$ (dashed line).}
\label{fig:rhobound1}
\end{center}
\end{figure}

\subsection{Direct bounds on charged heavy vectors} 

The coupling of the heavy vector charged mass eigenstates to a species of SM fermion takes the following form:
\begin{equation}
{\mathcal L}_{\rho ff} =  \rho^\pm_\mu \bar{f}^i_{SM} \gamma^\mu f^j_{SM} \left( \frac{g_\rho}{\sqrt{2}} c_\rho^\pm s_f^2 - \frac{g}{\sqrt{2}} s_\rho^\pm c_f^2 \right) \,,
\end{equation}
where $s_\rho^\pm$ is the sine of the mixing angle between the custodial triplet and the $SU(2)_L$ gauge fields, and $s_f$ is the sine of the mixing angle that defines the admixture of a composite fermion in a SM fermion mass eigenstate.\footnote{It is in principle possible to arrange the mixing angles associated with the fermion mass eigenstates such that the light fermion mass eigenstates do not couple at all to the heavy vectors.  In fact, this arises in a natural way in warped extra dimensional models in which the 5D bulk mass parameter for the fermions is close to the value which generates a flat profile for the light fermion.  The small coupling arises then from a wave-function orthogonality relation~\cite{Cacciapaglia:2004rb,Cacciapaglia:2006gp}.}  These couplings arise after diagonalizing the gauge boson and fermion mass matrices.  The mixing terms between fundamental and composite fermions 
follow from Eq.~(\ref{mixing}) in Appendix~\ref{sec:Appferm}.
The generic phenomenological effect of 
this partial compositeness
 will be to increase the relative branching fraction of the $\rho$ mesons to LH  fermion fields, decreasing the di-boson signal.  

The strongest direct search bounds arise from di-boson decays of the electrically charged $\rho$ mesons.  For most values of the parameters, there is a substantial branching fraction for the decay $\rho^\pm \rightarrow W^\pm Z$.  This is especially true when all SM fermions are fundamental, in which case this decay dominates the branching fraction for most values of the masses.  The final state which is most sensitive is the fully leptonic ``golden" channel in which $W^\pm Z \rightarrow 3 l + \nu$, which is experimentally very clean.  

In Fig.~\ref{fig:BRah0}, we show the branching fractions of the charged $\rho$'s to $W^\pm Z$, light quarks, third generation quarks, and leptons.  In the case that the SM fermions are all fundamental, the couplings to the $\rho$ mesons occur only through mixing with the SM gauge bosons (the fermions carry no charge under the $SU(2)_C$ gauge group).  In addition, the branching fractions are insensitive to the value of $a_\rho$.  This is shown in the first plot in Fig.~\ref{fig:BRah0}.  In some cases, the fermions (in particular the third generation) may be primarily composite fields.  In this case, they are expected to directly carry charge under $SU(2)_C$, with coupling strength $g_\rho$.   In the second plot in Fig~\ref{fig:BRah0}, we see that the decay width to third generation quarks dominates, however the branching fraction to SM gauge fields remains ${\mathcal O}(10\%)$, even in this extreme case of complete compositeness.

\begin{figure}[t]
\begin{center}
\includegraphics[scale=0.65]{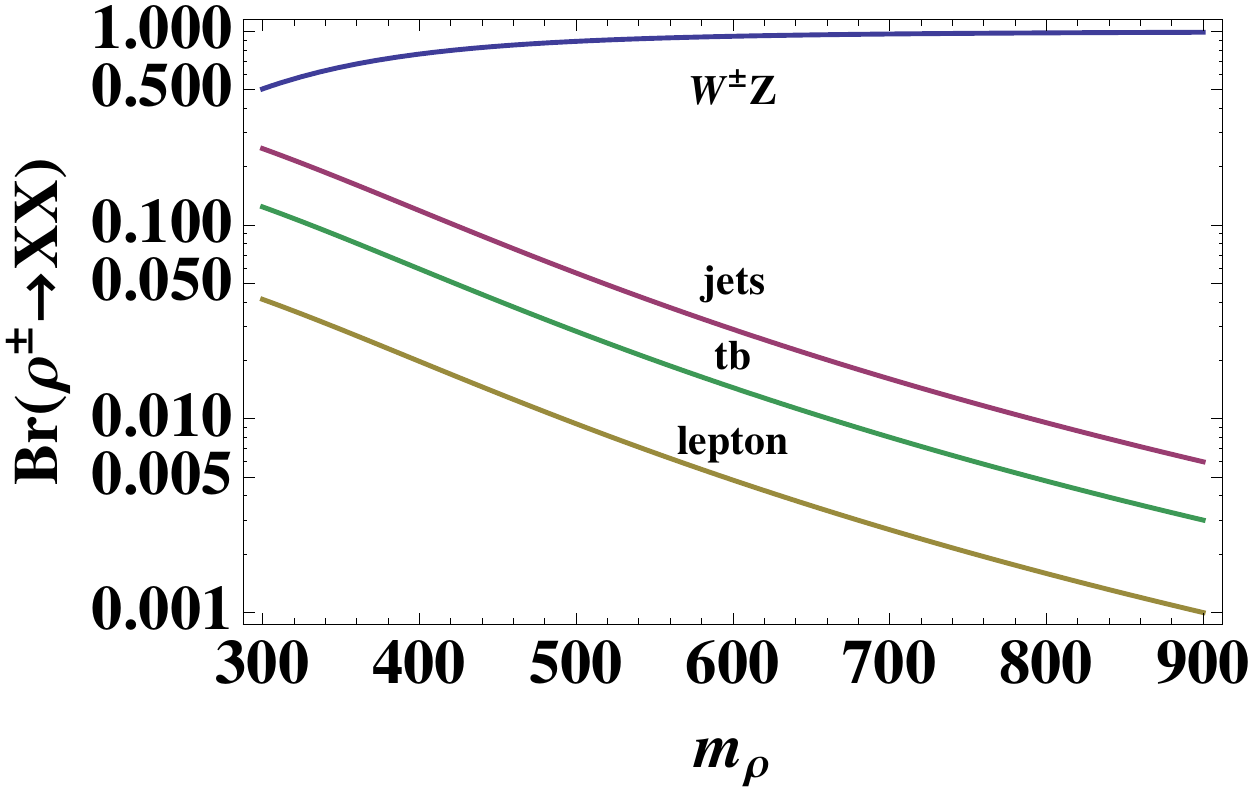}\includegraphics[scale=0.65]{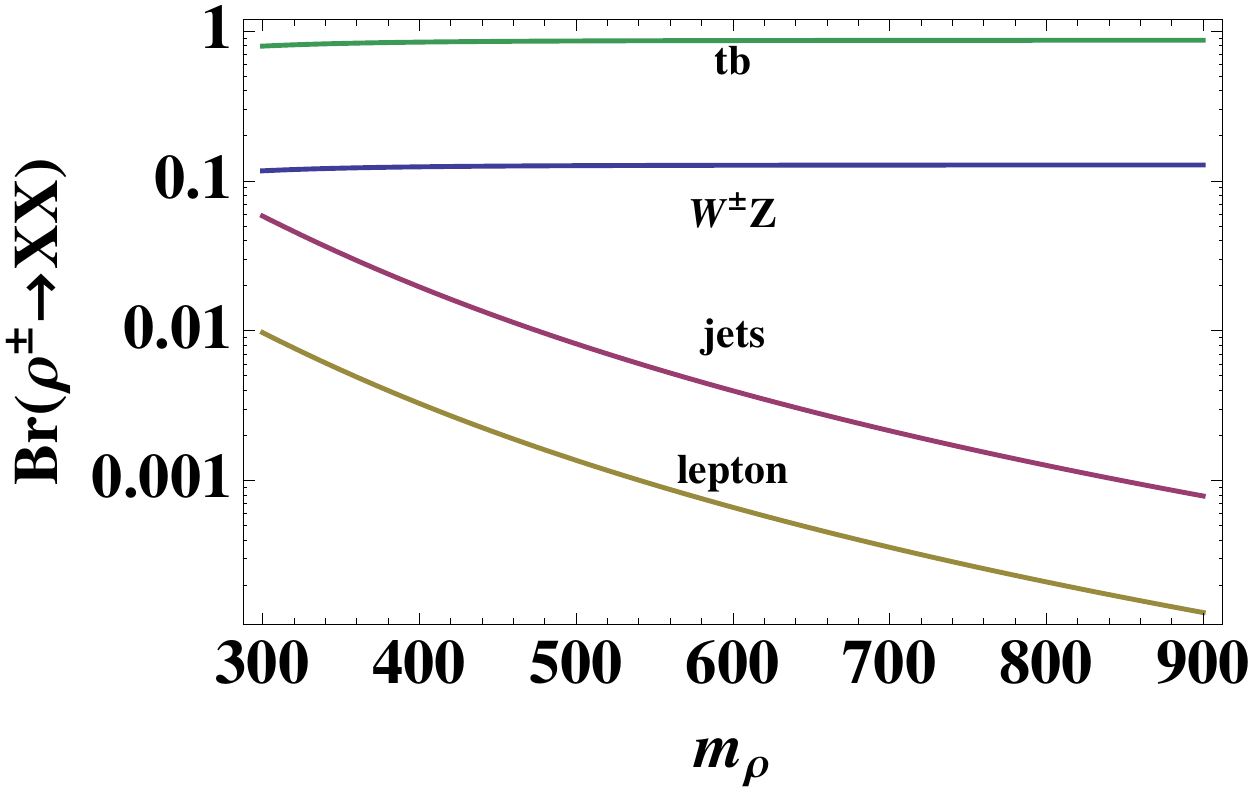}
\includegraphics[scale=0.65]{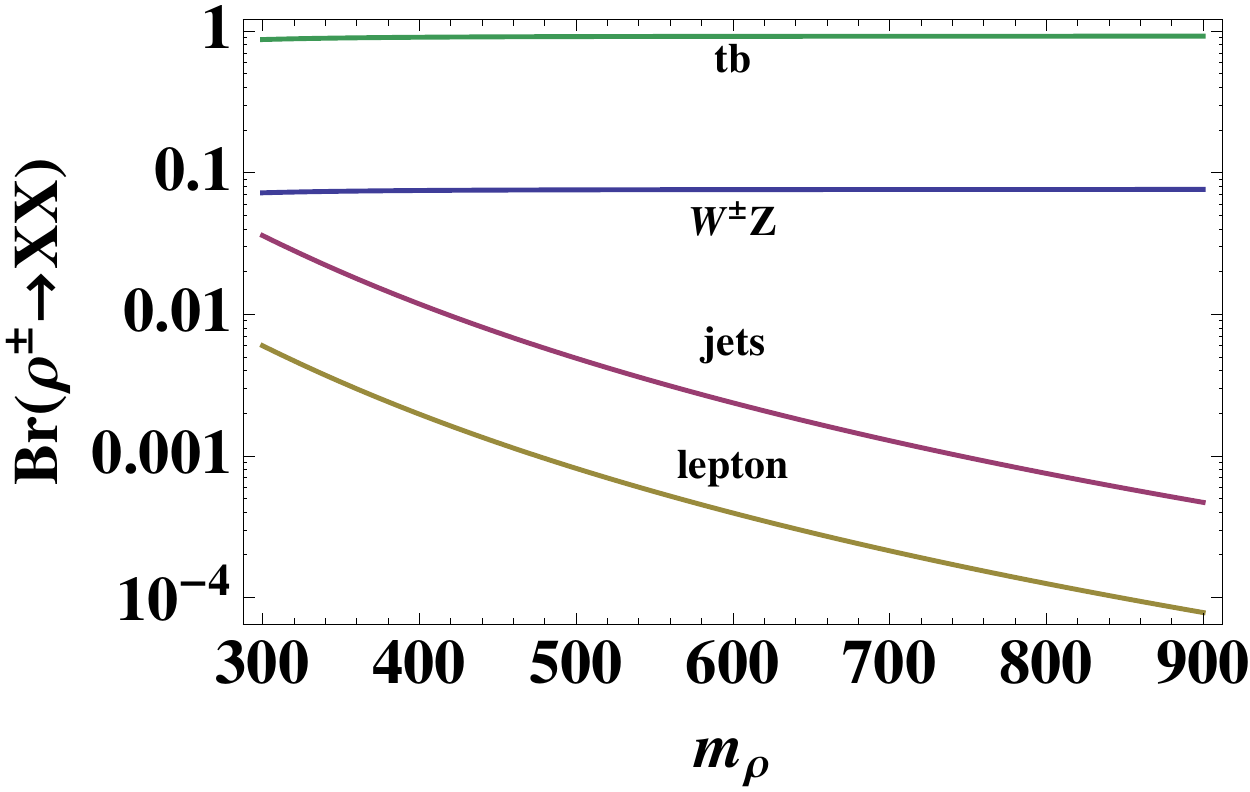}\includegraphics[scale=0.65]{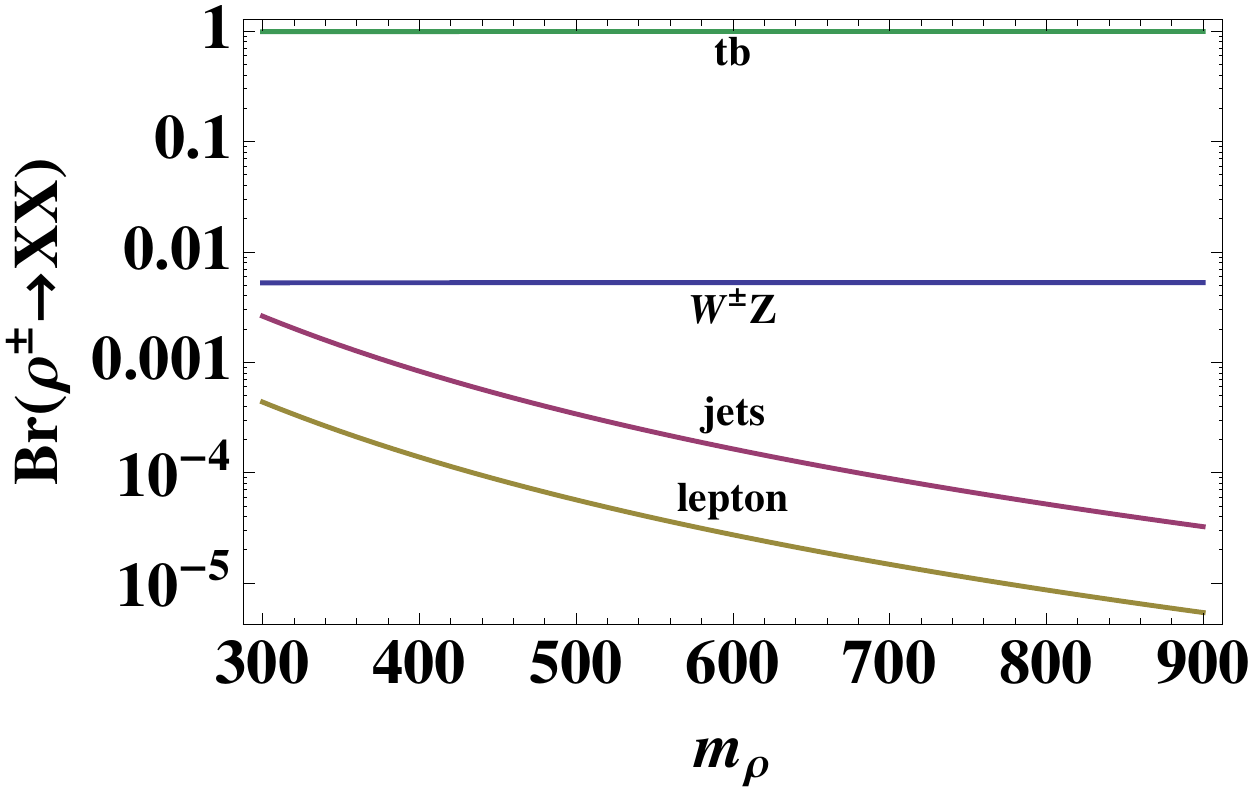}
\caption{The branching fractions of the charged $\rho$ vectors to $W^\pm Z$, light jets, $\bar{t}b$, and a single lepton species for different values of $a$, and for different degrees of compositeness for the LH third generation quarks.  On the top left is the case when all fermions are purely fundamental (the branching ratios are independent of $a$ in this case).  On the remaining plots, we assume that the third generation LH quarks are purely composite.  
The top right is composite third generation with $a=0$, the bottom left is $a=0.5$ and the bottom right is $a=0.9$. }
\label{fig:BRah0}
\end{center}
\end{figure}

With the dual requirements of satisfying the direct search constraints and of vector boson scattering processes remaining perturbative up to some cutoff scale $\Lambda$, 
we can explore the allowed parameter space of strongly coupled theories that contain a light Higgs and rho.  Both the perturbative unitarity constraints and the collider bounds depend on two parameters, $a$  and $m_\rho$.  For smaller values of $a$, the $\rho$'s must come in earlier to satisfy the perturbativity constraint, placing an upper bound on their masses.  However,  there is a tension since LHC searches place a lower bound on these masses. The production cross section for the $\rho^\pm$ at the LHC is shown in Fig.~\ref{fig:rhobound1}.

To constrain the model, we consider two scenarios, one in which the third generation fermions are completely fundamental, in which case the branching fraction to the $W^\pm Z$ final state dominates, and the other where the LH $t$ and $b$ fermions are completely composite, as described above.  These limits are shown in Fig.~\ref{fig:rhobound}.  In the regions on the left, which are bounded by solid lines, we show the direct collider exclusion bounds for these two scenarios.  The light grey region shows the constraints when the 3rd generation fermions are assumed to be completely elementary.  The large branching fraction to $W^\pm Z$, leads to a larger signal, and thus stronger bounds.  The darker grey region shows the constraints when the 3rd generation fermions are assumed to be completely composite (and thus have a direct coupling to the $\rho$'s).  The effect of this coupling on the phenomenology is to reduce the branching fraction to $W^\pm Z$, which results in a suppression of the $3l+\nu$ signal, and weaker constraints on $m_\rho$.  There is a large region of parameter space for the $\rho$'s that is consistent with direct search constraints, electroweak precision, and perturbative unitarity.  In particular, the region where $a$ is close to one is particularly favored both in terms of viable parameter space, and in terms of consistency with the current hints of a Higgs-like resonance near $125$~GeV.
\begin{figure}[ht]
\begin{center}
\includegraphics[height=3.3in]{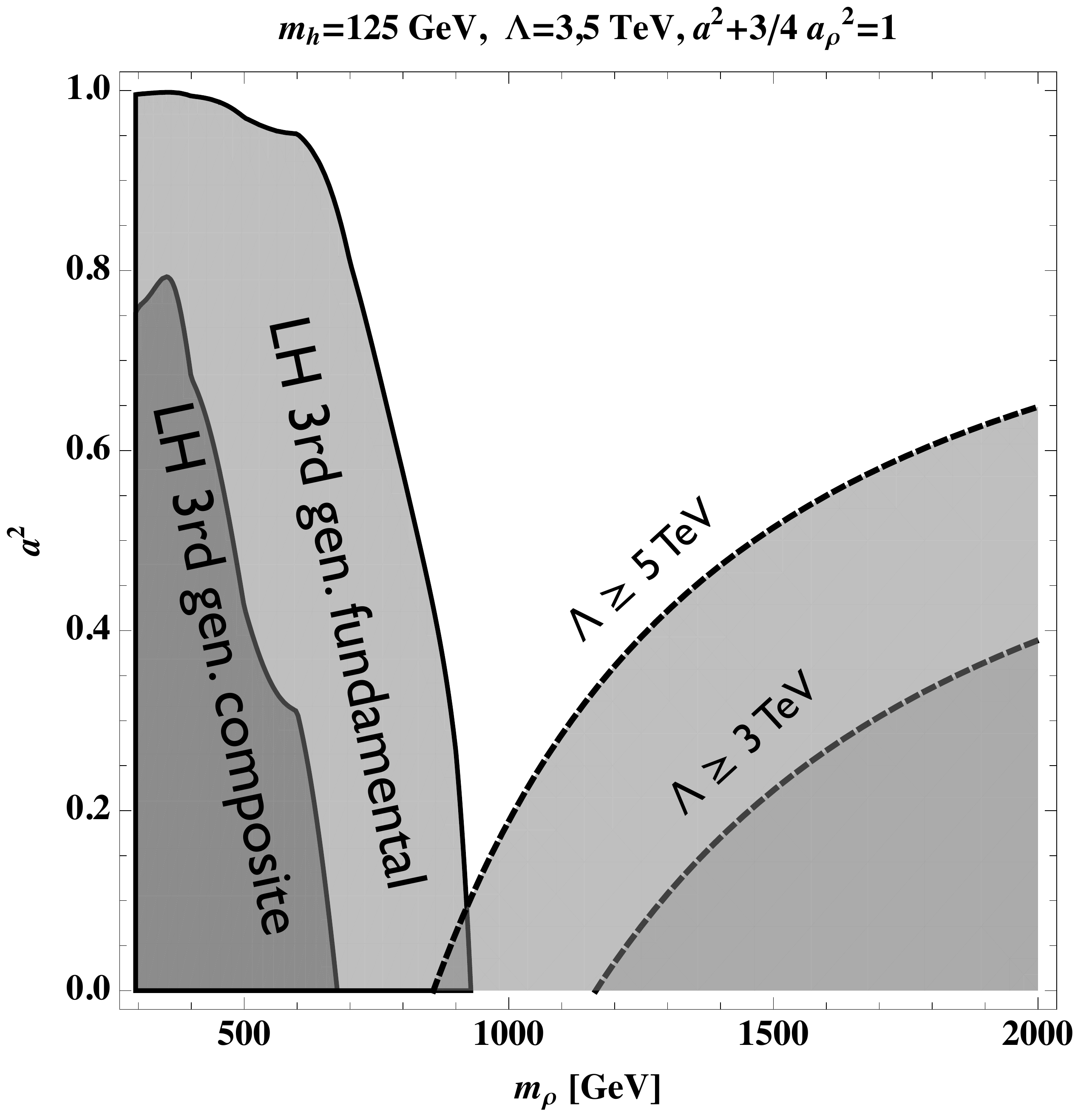}\includegraphics[height=3.3in]{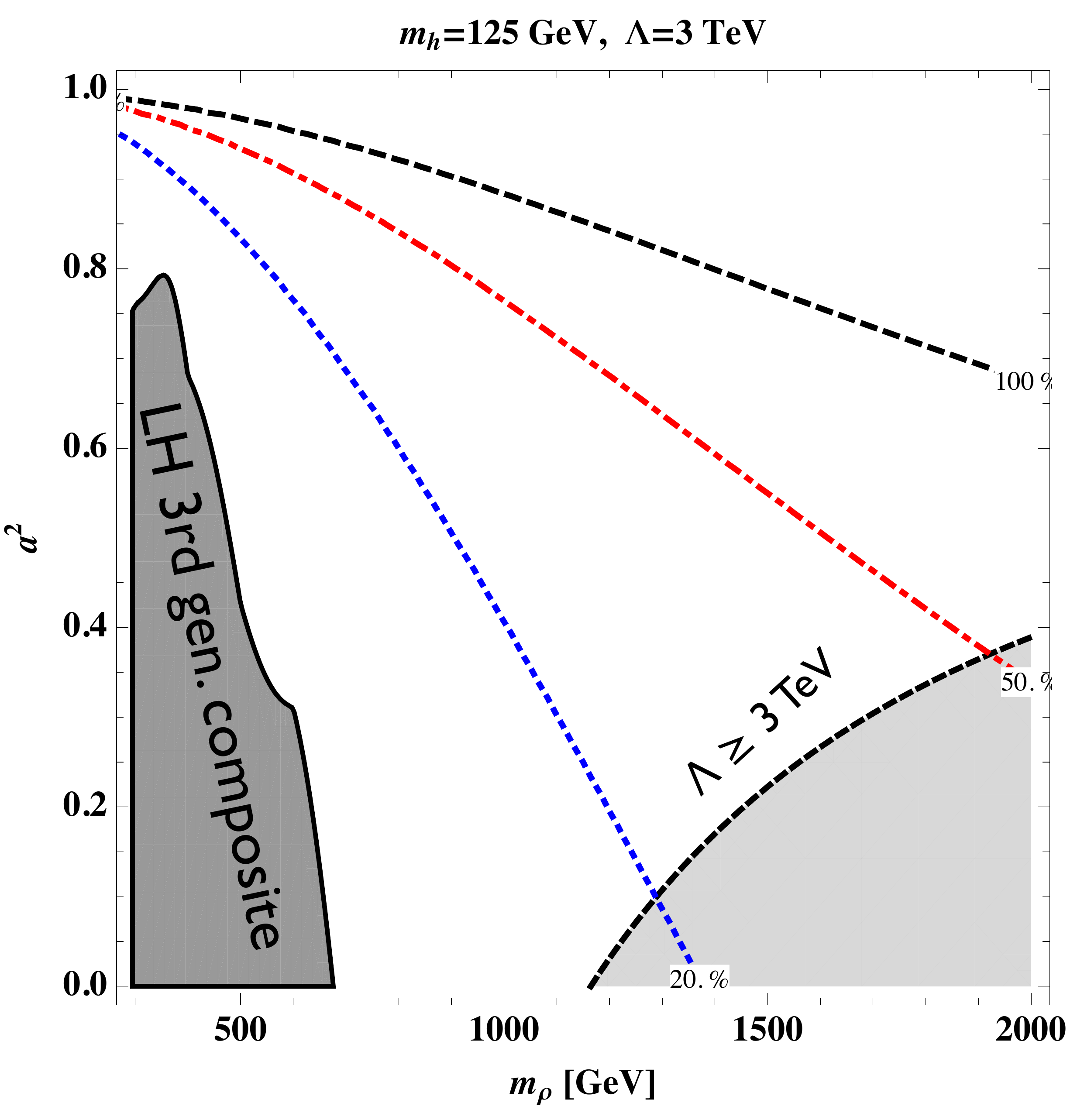}
\caption{In the plot on the left, we show the exclusion regions both from theoretical considerations (the dashed contours) arising from constraining the model to remain unitary up to the cutoff scale $\Lambda =3$ TeV or $\Lambda =5$ TeV, and from the CMS direct search constraints (solid contours).  Two extremes of compositeness for the third generation LH fermions are considered.  All fermions are presumed to be completely elementary for the first contour (excluded region shaded light gray), and in the other, the third generation quarks are taken to be purely composite (shaded dark gray) with fundamental 1st and 2nd generation fermions.  In the plot on the right, we superimpose the tuning required to satisfy constraints on the $S$-parameter on the unitarity bound on $m_\rho$ with $\Lambda = 3$~TeV, and on the collider bounds when the 3rd generation left-handed fermions are completely composite.}
\label{fig:rhobound}
\end{center}
\end{figure}

\subsection{Direct Bounds on Neutral Heavy Vectors}

The neutral $\rho$ mixes with the $Z$ and $\gamma$, and thus inherits couplings to fundamental fermions.  This mixing also leads to a $\rho^0 W^+ W^-$ coupling, which opens up the decay channel $\rho^0 \rightarrow W^+ W^-$.   A recent paper attempts to use the CMS and ATLAS Higgs search constraints to place limits on such particles~\cite{Eboli:2011ye}.  In principle, the Higgs searches in the $W^+W^-$ final state places limits on the $\rho^0$ mass, however this search is highly optimized for a Higgs boson with SM couplings, and the results are difficult to interpret in terms of a generic resonance search.  While the $\rho^0$ is primarily produced via Drell-Yan, the SM Higgs is produced in a combination of gluon and vector-boson-fusion, with VBF dominating in the high mass region.  

Additionally, if the third generation LH fermions are primarily composite, the $\rho^0$ may couple strongly to $\bar{t}t$ and $\bar{b} b$ final states, although this coupling is model dependent.  In fact, as discussed in Section~\ref{sec:phenolag}, the coupling of the $\rho^3$ to the composite top is 
suppressed
in the presence of a $P_{LR}$ symmetry.    Since the $\rho^0$ is in the $SU(2)_C$ triplet, and RH composite fermions are chosen to be singlets of this group, there are no large couplings of the $\rho^0$ to these composite degrees of freedom.  Extensive searches have been performed for $\bar{t}t$ resonances in the context of $Z'$ models and in searches for Kaluza-Klein gluons, and these can be used to place limits on the $\rho^0$ mass and couplings.

The $\rho^0$ is produced in Drell-Yan, like the $\rho^\pm$, and its cross section is similar in magnitude.  At $m_{\rho^0}\sim1$ TeV, the production cross section is about $.05$ pb.  This cross section is far below the current limits in either the boosted or non-boosted $\bar{t}t$ resonance searches, which are  in the $1-0.1$ pb range for $m_{\rho^0}\sim 1-3$ TeV.  The search for the charged components of the custodial triplet of vectors thus places the strongest limits, even when decays to composite $t$ and $b$ quarks dominate the width.

\section{Conclusions}

The LHC seems to be hinting at a 125 GeV Higgs boson. This still leaves open the possibility that this particle is a light resonance produced by strong interactions. In this case large deviations of the Higgs couplings to SM fields would be expected, which raises the question whether the theory remains perturbative up to the scale of the strong dynamics, $\Lambda$. In particular, longitudinal gauge boson scattering amplitudes may violate unitarity below the compositeness scale, which would imply that more resonances have to be below $\Lambda$. The concrete example we have investigated in this paper is when an additional custodial triplet $\rho^{\pm,0}$ is lighter than $\Lambda$, and whose couplings are close to the unitarity sum rules imposed by elastic and inelastic gauge boson and Higgs scattering amplitudes. We have found that such fields can significantly enhance the LHC rate for $h\to \gamma\gamma$, even when the Higgs couplings to the $W$, and $Z$ are suppressed. The most useful LHC channel to search for the $\rho$'s are the di-boson final states, in particular the $\rho^\pm \to W^\pm Z\to 3l+\nu$ channel. We reinterpreted the CMS search for $WZ$ final states and found that the bulk of the interesting parameter range 700 GeV$\lesssim m_\rho \lesssim$ 2 TeV is allowed as long as the tree-level Higgs couplings are not too far from their SM values. The $\rho^\pm \to W^\pm Z$ channel is the dominant decay channel if all fermions are elementary, and the second most important (and still quite significant) when the third generation is composite. The neutral $\rho^0$ can be searched for in the $W^+W^-$ channel, but a reinterpretation of the $W^+W^-$ Higgs bounds is quite challenging. On the other hand the cross section is currently too small to reach the sensitivity of the $t\bar{t}$ resonance searches.





\section*{Acknowledgements}

We thank Matthew Reece, Andrea Wulzer, and Riccardo Torre for useful conversations during the course of this work.  
J.H. thanks Cornell University for hospitality during the course of this work. 
B.B. is supported by the ERC Advanced Grant no.267985, ``Electroweak Symmetry Breaking, Flavour and Dark Matter: One Solution for Three Mysteries''} (\textit{DaMeSyFla}). C.C. and J.S. are supported in part by the NSF grant PHY-0757868.  J.H. was supported by the DOE under grant number DE-FG02-85ER40237.  J.T. was supported by the
Department of Energy under grant DE-FG02-91ER406746.


\appendix

\section*{Appendix}

\section{CCWZ}
\label{CCWZ}
\setcounter{equation}{0}
\setcounter{footnote}{0}

\subsection{The Goldstone-boson Lagrangian}

We review here the recipe to build effective field theories that involve NGB's associated to the spontaneous symmetry breaking pattern $\mathcal{G}/\mathcal{H}$. The NGB fields are associated with the coset space, i.e. they provide a map from spacetime to a group element $g(x)$ up to the equivalence relation $g(x)\sim g(x)h(x)$ where $h(x)\in\mathcal{H}$.
 It is then very convenient to use the CCWZ formalism \cite{Coleman:1969sm} which works with fields $\Pi(x) = \Pi^{\hat a}(x) T^{\hat a}$ that provide natural coordinates on the coset space around the identity, $g=U(\Pi)h$ where $U=\mbox{exp}(i\Pi)$. $T^{\hat a}$ represent the broken generators. The group transformation $U(\Pi)\rightarrow g_0 U(\Pi)=U(\Pi^\prime)h(\Pi,g_0)$ 
defines the action of $\mathcal{G}$ on $\Pi$
\begin{equation}
\label{Pi_tranf}
U(\Pi)\rightarrow U(\Pi^\prime)=g_0 U(\Pi)h^{-1}(\Pi,g_0)\,,
\end{equation}
that is a linear representation of $\mathcal{H}$ and non-linear, in general, for the other elements in $\mathcal{G}$.
Taking $\mathcal{G}$ as a symmetry of the UV dynamics then $\Pi$ enters in the effective Lagrangian only through its derivatives, coming from
\beq
-iU^{-1}\partial_\mu U=\Pi^{\hat a}_\mu T^{\hat a}+ E^{a}_\mu T^{a}\equiv
\Pi_\mu+ E_\mu \,,
\eeq
which decomposes along the broken and unbroken directions (because it is an element of the algebra of $\mathcal{G}$),
\bea
\Pi_\mu^{\hat a}=-i\text{Tr}[T^{\hat a} U^{-1}\partial_{\mu} U] \, , \qquad
E_\mu^a=-i\text{Tr}[T^a U^{-1}\partial_{\mu} U].
\eea
These have different transformation rules under (\ref{Pi_tranf}):
\bea
\label{Pi_transf}
\Pi_\mu \!\!\!&\rightarrow&\!\!\! h(\Pi,g_0)\Pi_\mu h^{-1}(\Pi,g_0) \\
 E_\mu \!\!\!&\rightarrow&\!\!\! h(\Pi,g_0)E_\mu h^{-1}(\Pi,g_0)-i h(\Pi,g_0)\partial h^{-1}(\Pi,g_0)\, .
\eea
In particular, $E_\mu(\Pi)$ transforms as a gauge field under $\mathcal{H}$ and one can define a
covariant derivative $\nabla_{\mu}=\partial_\mu +i  E_\mu$ to couple $U(\Pi)$ to light  matter fields $\psi$ that provide linear representations $D$ of $\mathcal{H}$, $\psi\rightarrow D(h(\Pi))\psi$. 

When $\mathcal{H}$ is symmetric  there exists an automorphism $R$ of the algebra that changes the sign of the broken generators
\beq
T^{\hat a} \to -T^{\hat a}\, , \qquad T^a \to T^a \, ,
\eeq
that is $\Pi \to - \Pi$, $\Pi_\mu^{\hat a} \to - \Pi_\mu^{\hat a}$ and $E_\mu^{a} \to + E_\mu^{a}$.
In such a case the leading order expressions for $\Pi_\mu$ and $E_\mu$ are simply
\begin{equation}
\Pi_\mu=\partial_\mu \Pi-\frac{1}{6}[\Pi,[\Pi,\partial_\mu \Pi]]+\mathcal{O}(\Pi^5)\, ,  \qquad E_\mu=-\frac{i}{2}[\Pi,\partial_\mu \Pi]+\mathcal{O}(\Pi^4)\,.
\end{equation}
Note also that under $R$  one has
\begin{equation}
g_0 U(\Pi)=U(\Pi^\prime)h(\Pi, g_0)\longrightarrow R(g_0)U(\Pi)^{-1}=U(\Pi^\prime)^{-1}h(\Pi,g_0) \,  ,
\end{equation}
so that eliminating $h(\Pi,g_0)$ the transformation is linear on $\Sigma=U^2(\Pi)$
\begin{equation}
\Sigma \rightarrow g_0 \Sigma(\Pi) R(g_0)^{-1} \,.
\end{equation}

Given the transformation rule (\ref{Pi_transf}) we can write the lowest order (in fields and derivatives) Lagrangian
\beq
\mathcal{L}^{(2)}_{\Pi}= \frac{v^2}{2} (\Pi_\mu^{\hat a})^2=\frac{1}{2}(\partial_\mu\pi )^2+\ldots \qquad \pi=v \Pi\,.
\eeq
which is invariant under the full group $\mathcal{G}$.

 Finally, when $\mathcal{H}^\prime \subseteq \mathcal{G}$ is weakly gauged we just need to introduce  the usual covariant derivatives, $\partial_\mu \rightarrow D_\mu = \partial_\mu - iA_\mu$, and add the kinetic terms for $A_\mu$.

In this paper we have $\mathcal{G}=SU(2)_L\times SU(2)_R$ broken to the diagonal $SU(2)_{C=L+R}$. We also have an automorphism $P_{LR}$ that exchanges $L \leftrightarrow R$  making $O(4)/O(3)$ the actual symmetry breaking pattern.
Thanks to parity it is actually easier to work with the standard bi-doublet $U^2=\Sigma=(\mathbf{2},\mathbf{2})$ that transforms linearly 
\begin{equation}
\Sigma \rightarrow L\Sigma R^\dagger\,.
\end{equation}
The inclusion of the SM gauge fields is now trivially achieved by the replacement $\partial_\mu\Sigma \rightarrow D_\mu\Sigma$ where $D_\mu=\partial_\mu+igT_L+ig^\prime T^3_R$ in the action for $\Sigma$
\beq
\mathcal{L}^{(2)}_{\pi}= \frac{v^2}{4}\mbox{Tr }[D_\mu\Sigma^\dagger D_\mu\Sigma] \, .
\eeq

\subsection{Light matter fields}

Light matter fields $\psi$ transform as linear representations of the unbroken group $\mathcal{H}$. Their couplings to NGB's come from the  connection $E_\mu(\Pi)$ that defines an effective covariant derivative $\nabla_\mu=\partial_\mu+iE_\mu$.
For example, fermions in the fundamental have 
$$
\mathcal{L}=\bar{\psi}i\slashed{\nabla}\psi=f^{abc} \pi^b \partial_\mu \pi^c \bar{\psi}T^{a} \gamma^\mu \psi+\ldots
$$

\subsection{Light  vectors}

As discussed in Section~\ref{eff_rho} it is often convenient to introduce a light vector that transforms as a gauge boson, namely
\beq
\rho_\mu \to  h^{-1}\rho_\mu h+i h^{-1}\partial_\mu h\,.
\label{gauge}
\eeq
Its Lagrangian at the lowest order, again in fields and derivatives, is
\beq
\mathcal{L}^{(2)}_{\rho}=- \frac{1}{4} (\rho_{\mu \nu}^a)^2 + a_\rho^2 \frac{v^2}{2} \left( g_\rho \rho_\mu^a - E_\mu^a \right)^2 \,,
\label{rho_L}
\eeq
where 
$\rho_{\mu\nu}\equiv \partial_{\mu}\rho_{\nu}-\partial_{\nu}\rho_{\mu}+i[\rho_\mu,\rho_\nu]$. 
Besides the mass of the rho, $m_\rho=a_\rho v g_\rho$, and the coupling to  NGB's, 
the second term in \eq{rho_L} generates a new $\pi^4$ vertex that is relevant in $\pi\pi$ elastic scattering.

\subsection{Singlet scalars}

A singlet under $\mathcal{H}$ couples to all invariant operators that one can build with the other fields (such as $\pi$, $\rho$ and matter fields) and that are allowed by the discrete symmetries. In our model with $\mathcal{H}=SU(2)_C$
the Higgs boson particle $h$ is even under $P_{LR}$ and thus
\beq
\label{scalar_coup}
\mathcal{L}^{(2)}_{h}=\frac{1}{2} (\partial_\mu h)^2 + V(h) + \frac{v^2}{2} \left( 2 a_h \frac{h}{v} + b_h \frac{h^2}{v^2} \right) (\Pi_\mu^{\hat a})^2 + \frac{v^2}{2} \left( 2 c_h \frac{h}{v} + d_h \frac{h^2}{v^2}  \right) \left( g_\rho \rho_\mu^a - E_\mu^a \right)^2 \, .
\eeq
A $P_{LR}$-odd scalar $H$ has instead the Lagrangian
\beq
\mathcal{L}^{(2)}_{H}=\frac{v^2}{2} \left( 2 a_H \frac{H}{v} + 2 b_H \frac{h H}{v^2} \right) \Pi_\mu^{\hat a} \left( g_\rho \rho^{\mu a} - E^{\mu a} \right) \delta^{\hat a a} 
+ \frac{v^2}{2} c_H \frac{H^2}{v^2} (\Pi_\mu^{\hat a})^2
+ \frac{v^2}{2} d_H \frac{H^2}{v^2} \left( g_\rho \rho_\mu^a - E_\mu^a \right)^2 \, .
\eeq

\subsection{Composite fermions\label{sec:Appferm}}

The couplings of elementary fermions to the strong sector are given by
\beq
{\cal L}_{\mathrm{mix}}= 
		(\bar u_L, \bar d_L) \epsilon_L^{A} \mathcal{Q}_{A} 	+
		\bar u_R \epsilon_R^{B} \, \mathcal{U}_B +
		h.c.\, ,
\label{mixing}
\eeq
where, as explained in Section~\ref{eff_rho}, we take the composite operators to transform as
 $\mathcal{Q} \sim \mathbf{(2,2)_{2/3}}$ and $\mathcal{U} \sim \mathbf{(1,1)_{2/3}}$ of $SU(2)_L \times SU(2)_R \times U(1)_X$, with $A=1,2,3,4$ a $SO(4)$ index. $\epsilon_{L,R}$ are a set of couplings which parametrize the degree of compositeness.

The couplings of the chiral fermion $\mathcal{Q}_A$ to NGB's and to the $SU(2)_C$ gauge vector $\rho$ is obtained from the CCWZ formalism.
The low-energy Lagrangian is determined by $SU(2)_C$ symmetry.
Therefore we must decompose the $\mathcal{Q}$ multiplet, transforming as a \textbf{4} of $SO(4)$, into $SO(3)$ representations, that is a \textbf{1}, $\eta = \mathcal{Q}_A (U^{A}_4)^*$, and a \textbf{3}, $\psi_a = \mathcal{Q}_A (U^{A}_a)^*$ ($a = 1, 2, 3$), 
where $U$ is the NGB matrix.
Then, the $SO(3)$ invariants involving the $\rho$ are
\bea
{\cal O}_{+} \!\!\! &=& \!\!\! \bar \psi_a \gamma^\mu (g_\rho \rho_\mu - E_\mu)_b \psi_c \epsilon_{abc} \\
{\cal O}_{-} \!\!\! &=& \!\!\! \bar \psi_a \gamma^\mu \eta (g_\rho \rho_\mu - E_\mu)_b \delta_{ab} \,.
\eea
At leading order in the NGB's, $U = 1$, the invariants read,
\bea
{\cal O}_+ &=& \frac{g_\rho}{2} \rho_\mu^0 (\bar x_L \gamma^\mu x_L-\bar d_L \gamma^\mu d_L) + \frac{g_\rho}{2 \sqrt 2} \rho^+ (\bar x_L \gamma^\mu u_L - \bar x_L \gamma^\mu u'_L + \bar{u}'_L \gamma^\mu d_L - \bar u_L \gamma^\mu d_L) + h.c. \nonumber \\
{\cal O}_- &=& \frac{g_\rho}{2} \rho_\mu^0 (\bar u'_L \gamma^\mu u'_L - \bar u_L \gamma^\mu u_L) - \frac{g_\rho}{2 \sqrt 2} \rho^+  (\bar x_L \gamma^\mu u_L + \bar x_L \gamma^\mu u'_L + \bar{u}'_L \gamma^\mu d_L + \bar u_L \gamma^\mu d_L) + h.c. \,, \nonumber \\
\label{eq:compositefraction}
\eea
where we have decomposed the $\mathcal{Q}$ multiplet in components,
\beq
\mathcal{Q} = \left( \begin{array}{cc} u_L  & x_L \\ d_L & u'_L \end{array} \right) = 1 \eta + i \sigma^a \psi_a.
\eeq
$x_L$ has electric charge $5/3$ and $u'_L$ has $2/3$.
Note that $x_L$ and $u'_L$ do not have to be light compared to the cut-off $\Lambda$, so we do not have to include them at all, and that if $u_L$ and $d_L$ are not totally composite, they should be rescaled by the degree of compositeness, $\epsilon_L$.

Finally, notice that the two invariants have different transformation properties under $P_{LR}$.
$\mathcal{O}_+$ is even while $\mathcal{O}_-$ is odd.
One can easily check this by noticing that $P_{LR}[\rho,E] = + (\rho,E)$, while $P_{LR}[\mathcal{Q}] = + 1 \eta - i \sigma^a \psi_a$.

\section{Resonance decays into $VV$, $Vh$}
\label{rhoVh}
\setcounter{equation}{0}
\setcounter{footnote}{0}

In this appendix we elaborate on the $\rho$ decay width into gauge boson pairs $VV^\prime$ and $Vh$.
Because of custodial symmetry it is enough to focus on $\rho^0 \rightarrow WW$ and, by means of the equivalence theorem, we can look at the following Lagrangian 
\begin{equation}
\label{Lpipi_decay}
\mathcal{L}_{\rho \pi\pi}=\frac{1}{2}a_\rho^2 g_\rho \rho^3_\mu(\partial_\mu \pi^1 \pi^2-1\leftrightarrow2)=a_\rho^2 g_\rho \rho^3_\mu \partial_\mu \pi^1 \pi^2+\ldots
\end{equation}
where the ellipses contain terms that vanish on-shell.
The width is dominated by the longitudinal polarizations and is given by
\begin{equation}
\Gamma(\rho\rightarrow VV)\simeq \Gamma(\rho\rightarrow \pi\pi)=\frac{1}{192\pi}a_\rho^4 g_\rho^2 m_\rho\,.
\end{equation}

Let us compare this result  with the $\Gamma(\rho\rightarrow hV)$ that arises from
\begin{equation}
\mathcal{L}=\lambda v h V_\mu \rho_\mu\,.
\end{equation}
For example we have  $\lambda \simeq g^3/(4g_\rho)\times c$ where $c = a-c_\rho$ is an order one parameter
, see \eq{crhoV}. Also in this case  the width is dominated by the longitudinal polarizations $V_L \sim \partial\pi/m_V$ 
and therefore, comparing with (\ref{Lpipi_decay}), we see that
\begin{equation}
\frac{\Gamma(\rho\rightarrow h V)}{\Gamma(\rho\rightarrow V V)}\simeq 4c^2 \left(\frac{g}{2 a_\rho g_\rho}\right)^4=4c^2 \left(\frac{m_W}{m_\rho}\right)^4 \simeq 3 \cdot 10^{-4} \left( \frac{c}{1} \right)^2 \left( \frac{1 \TeV}{m_\rho} \right) \,.
\end{equation}
A ratio close to $1$ needs $c\sim 4\pi$ and $m_\rho\sim 400$ GeV.

Finally, we compare these decay rates when the Higgs is a pNGB and $P_{LR}$ is broken (or one replaces the $\rho$ with an axial vector $A$). One can add the following vertex (see Eq.~(34) in \cite{Contino:2011np})
\begin{equation}
\mathcal{L}=a_\rho^2 g_\rho h \partial_\mu \pi^k \rho^k_\mu+\ldots
\end{equation}
where again the ellipses contain terms that vanish on-shell. Comparing with (\ref{Lpipi_decay}) we see that
\begin{equation}
\frac{\Gamma(\rho\rightarrow h V)}{\Gamma(\rho\rightarrow V V)}= 1 \,,
\end{equation}
which reflects the fact that $h$ and $\pi^i$ fit in a $\mathbf{4}$ of $SO(4)$ above the EW scale $v$. In fact, models where the Higgs is not part of a  $\mathbf{4}$, like for a dilaton, typically give
\begin{equation}
\frac{\Gamma(\rho\rightarrow h V)}{\Gamma(\rho\rightarrow V V)}= \mathcal{O}(1) \,,
\end{equation}
still assuming no $P_{LR}$.


\bibliographystyle{utphys.bst}
\bibliography{doubleCHiggs}

\end{document}